\pdfoutput=1

\documentclass[11pt,twoside,a4paper,cmspaper,final,collab]{cms-tdr}
\def\svnVersion{450921P}\def\cmsCernNoTag{CERN-EP-2017-219}\def\cmsCernDate{\today}\def\cmsMessage{Published in the European Physical Journal C as \href{http://dx.doi.org/10.1140/epjc/s10052-018-5567-9}{\doi{10.1140/epjc/s10052-018-5567-9.}}}

\begin{document}\cmsNoteHeader{SMP-16-017}

\hyphenation{had-ron-i-za-tion}
\hyphenation{cal-or-i-me-ter}
\hyphenation{de-vices}
\RCS$Revision: 450777 $
\RCS$HeadURL: svn+ssh://svn.cern.ch/reps/tdr2/papers/SMP-16-017/trunk/SMP-16-017.tex $
\RCS$Id: SMP-16-017.tex 450777 2018-03-13 09:25:23Z asavin $
\newlength\cmsFigWidth
\ifthenelse{\boolean{cms@external}}{\setlength\cmsFigWidth{0.48\textwidth}}{\setlength\cmsFigWidth{0.75\textwidth}}
\ifthenelse{\boolean{cms@external}}{\providecommand{\cmsLeft}{upper\xspace}}{\providecommand{\cmsLeft}{left\xspace}}
\ifthenelse{\boolean{cms@external}}{\providecommand{\cmsRight}{lower\xspace}}{\providecommand{\cmsRight}{right\xspace}}
\ifthenelse{\boolean{cms@external}}{\providecommand{\CMSLeft}{Top\xspace}}{\providecommand{\CMSLeft}{Left\xspace}}
\ifthenelse{\boolean{cms@external}}{\providecommand{\CMSRight}{Bottom\xspace}}{\providecommand{\CMSRight}{Right\xspace}}
\providecommand{\ptvecl}{\ensuremath{\ptvec^{\kern0.2em\ell}\xspace}}
\newlength\cmsTabSkip
\setlength\cmsTabSkip{1.8ex}
\providecommand{\ZZ}{\ensuremath{\cPZ\cPZ}\xspace}
\providecommand{\pp}{\ensuremath{\Pp\Pp}\xspace}
\newcommand{\elfour}{\ensuremath{4\ell}\xspace}
\newcommand{\elel}{\ensuremath{2\ell}\xspace}
\newcommand{\elelpr}{\ensuremath{2\ell'}\xspace}
\newcommand{\elpelm}{\ensuremath{\ell^+\ell^-}\xspace}
\newcommand{\elpelmpr}{\ensuremath{\ell^{\prime+}\ell^{\prime-}}\xspace}
\newcommand{\elelelel}{\ensuremath{\elpelm\elpelmpr}\xspace}
\newcommand{\twoeltwoel}{\ensuremath{\elel\elelpr}\xspace}
\newcommand{\elfourtau}{\ensuremath{2\ell2\ell^{\prime\prime}}\xspace}
\newcommand{\eleltautau}{\ensuremath{\ell^+\ell^-\ell^{\prime\prime+}\ell^{\prime\prime-}}\xspace}
\newcommand{\eltau}{\ensuremath{2\ell2\Pgt}}

\cmsNoteHeader{SMP-16-017}
\title{Measurements of the $\Pp\Pp\to \Z\Z$ production cross section and the $\Z \to 4\ell$ branching fraction, and constraints on anomalous triple gauge couplings at $\sqrt{s} = 13\TeV$}
\titlerunning{$\pp \to \ZZ$, $\Z \to 4\ell$, and aTGC at $\sqrt{s}=13\TeV$}

\author{CMS Collaboration}

\date{\today}

\abstract{
Four-lepton production in proton-proton collisions, $\Pp\Pp \to (\Z / \gamma^*)(\Z /\gamma^*) \to 4\ell$, where $\ell = \Pe$ or $\mu$, is studied at a center-of-mass energy of 13\TeV with the CMS detector at the LHC. The data sample corresponds to an integrated luminosity of 35.9\fbinv. The ZZ production cross section, $\sigma(\Pp\Pp \to \Z\Z) = 17.2 \pm 0.5\stat\pm 0.7\syst\pm 0.4\thy\pm 0.4\lum\unit{pb}$, measured using events with two opposite-sign, same-flavor lepton pairs produced in the mass region $60 < m_{\ell^+\ell^-} < 120\GeV$, is consistent with standard model predictions. Differential cross sections are measured and are well described by the theoretical predictions. The Z boson branching fraction to four leptons is measured to be $\mathcal{B}(\Z \to 4\ell) = 4.83 _{-0.22}^{+0.23} \stat _{-0.29}^{+0.32} \syst \pm 0.08 \thy \pm 0.12 \lum \times 10^{-6}$ for events with a four-lepton invariant mass in the range $80 < m_{4\ell} < 100\GeV$ and a dilepton mass $m_{\ell\ell} > 4\GeV$ for all opposite-sign, same-flavor lepton pairs. The results agree with standard model predictions. The invariant mass distribution of the four-lepton system is used to set limits on anomalous ZZZ and ZZ$\gamma$ couplings at 95\% confidence level: $-0.0012<f_4^\Z<0.0010$, $-0.0010<f_5^\Z<0.0013$, $-0.0012<f_4^{\gamma}<0.0013$, $-0.0012<f_5^{\gamma}< 0.0013$.
}

\hypersetup{%
pdfauthor={CMS Collaboration},%
pdftitle={Measurements of the pp to ZZ production cross section and the Z to 4l branching fraction, and constraints on anomalous triple gauge couplings at sqrt(s) = 13 TeV},%
pdfsubject={CMS},%
pdfkeywords={CMS, physics, electroweak}}

\maketitle

\section{Introduction}

{\tolerance=800
Measurements of diboson production at the CERN LHC allow precision
tests of the standard model (SM).
In the SM, $\cPZ\cPZ$ production proceeds mainly through quark-antiquark
$t$- and $u$-channel scattering
diagrams. In calculations at higher orders in quantum chromodynamics (QCD),
gluon-gluon fusion also contributes via
box diagrams with quark loops. There are no tree-level contributions to
$\cPZ\cPZ$ production from triple gauge boson vertices in the SM.
Anomalous triple gauge couplings (aTGC) could be induced by
new physics models such as supersymmetry~\cite{Gounaris:2000tb}.
Nonzero aTGCs may be parametrized using an effective
Lagrangian as in Ref.~\cite{Hagiwara}.
In this formalism, two $\PZ\PZ\PZ$ and two $\PZ\PZ\gamma$ couplings are
allowed by electromagnetic gauge invariance and Lorentz invariance for on-shell
$\PZ$ bosons. These are described by two CP-violating ($f_4^{\mathrm{V}}$) and
two CP-conserving ($f_5^{\mathrm{V}}$) parameters,
where ${\mathrm{V}} = \PZ$ or $\gamma$.
\par}

{\tolerance=800
Previous measurements of the ZZ production cross section by the
CMS Collaboration were performed for pairs of on-shell $\PZ$ bosons, produced in the dilepton
mass range
60--120\GeV~\cite{Chatrchyan:2012sga, CMS:2014xja, Khachatryan:2015pba, Khachatryan:2016txa}.
These measurements were made with data sets corresponding to
integrated luminosities of 5.1\fbinv at $\sqrt{s} = 7\TeV$ and
19.6\fbinv at $\sqrt{s} = 8\TeV$ in
the $\cPZ\cPZ \to \elfourtau$ and $\cPZ\cPZ \to \elel 2\nu$
decay channels, where $\ell = \Pe$ or $\Pgm$ and
$\ell^{\prime\prime} = \Pe, \Pgm$, or $\Pgt$, and with an integrated luminosity of
2.6\fbinv at $\sqrt{s} = 13\TeV$ in the $\ZZ \to \twoeltwoel$ decay channel, where
$\ell' = \Pe$ or $\Pgm$. All of them agree with SM predictions.
The ATLAS Collaboration produced similar
results at $\sqrt{s} = 7$, 8, and 13\TeV~\cite{Aad:2012awa, Aad:2015rka, Aad:2015zqe, Aaboud:2017rwm},
which also agree with the SM.
These measurements are important for testing predictions that were
recently made available at next-to-next-to-leading order (NNLO) in
QCD~\cite{Cascioli:2014yka}. Comparing these predictions
with data at a range of center-of-mass energies provides information about
the electroweak gauge sector of the SM. Because the uncertainty of the CMS
measurement at $\sqrt{s} = 13\TeV$~\cite{Khachatryan:2016txa} was dominated by
the statistical uncertainty of the observed data, repeating and extending the
measurement with a larger sample of proton-proton collision data at
$\sqrt{s} = 13\TeV$ improves the precision of the results.

The most stringent previous limits on $\cPZ\cPZ\PZ$ and $\cPZ\cPZ\gamma$ aTGCs
from CMS were set using the 7~and $8\TeV$ data samples:
$-0.0022<f_4^\cPZ<0.0026$, $-0.0023<f_5^\cPZ<0.0023$, $-0.0029<f_4^{\gamma}<0.0026$, and
$-0.0026<f_5^{\gamma}<0.0027$ at 95\% confidence level (CL)~\cite{Khachatryan:2015pba,CMS:2014xja}.
Similar limits were obtained by the ATLAS Collaboration~\cite{Aaboud:2016urj}, who also
recently produced limits using $13\TeV$ data~\cite{Aaboud:2017rwm}.
\par}

Extending the dilepton mass range to lower values
allows measurements of
$\left(\cPZ/\gamma^{\ast}\right) \left(\cPZ/\gamma^{\ast}\right)$ production,
where $\cPZ$ indicates an on-shell $\cPZ$ boson or an off-shell $\cPZ^{\ast}$ boson.
The resulting sample includes Higgs boson events in the
$\PH  \to  \cPZ\cPZ^{\ast}  \to  \twoeltwoel$ channel,
and rare decays of a single $\cPZ$ boson to four leptons. The
$\cPZ  \to  \ell^+\ell^-\gamma^{\ast}  \to  2\ell2\ell'$
decay was studied in detail at LEP~\cite{Buskulic:1994gk} and was
observed in pp collisions by CMS~\cite{CMS:2012bw, Khachatryan:2016txa} and
ATLAS~\cite{Aad:2014wra}. Although the branching fraction for this decay is orders of
magnitude smaller than that for the {$\cPZ \to \ell^+\ell^-$} decay, the
precisely known mass of the $\cPZ$ boson makes the four-lepton mode
useful for calibrating mass measurements of the nearby Higgs boson
resonance.

This paper reports a study of four-lepton production
($\pp  \to  \twoeltwoel$, where $2\ell$ and $2\ell'$ indicate
opposite-sign pairs of electrons or muons) at $\sqrt{s} = 13\TeV$ with a data set corresponding to
an integrated luminosity of $35.9 \pm 0.9$\fbinv recorded in 2016. Cross
sections are measured for nonresonant production of pairs of $\cPZ$ bosons,
$\pp  \to  \cPZ\cPZ$, where both $\cPZ$ bosons are produced on-shell,
defined as the mass range 60--120\GeV, and resonant $\pp  \to  \cPZ
\to \elfour$ production. Detailed discussion of resonant Higgs boson
production decaying to $\cPZ\cPZ^\ast$, is beyond the scope of this paper
and may be found in Ref.~\cite{Sirunyan:2017exp}.

\section{The CMS detector}

A detailed description
of the CMS detector, together with a definition of the coordinate system used
and the relevant kinematic variables, can be found
in Ref.~\cite{Chatrchyan:2008zzk}.

The central feature of the CMS apparatus is a superconducting solenoid of
6\unit{m} internal diameter, providing a magnetic field of 3.8\unit{T}.
Within the solenoid volume are a silicon pixel and strip
tracker, a lead tungstate crystal electromagnetic calorimeter (ECAL), and a
brass and scintillator hadron calorimeter,
which provide coverage in pseudorapidity $\abs{ \eta } < 1.479 $ in a cylindrical barrel
and $1.479 < \abs{ \eta } < 3.0$ in two endcap regions.
Forward calorimeters extend the
coverage provided by the barrel and
endcap detectors to $\abs {\eta} < 5.0$. Muons are measured in gas-ionization detectors embedded in
the steel flux-return yoke outside the solenoid
in the range $\abs{\eta} < 2.4$, with
detection planes made using three technologies: drift tubes, cathode strip
chambers, and resistive plate chambers.

Electron momenta are estimated by combining energy measurements in the
ECAL with momentum measurements in the tracker. The momentum resolution for
electrons with transverse momentum $\pt \approx 45\GeV$
from $\Z \to \Pep \Pem$ decays
ranges from 1.7\% for nonshowering electrons in the barrel region to 4.5\% for
showering electrons in the endcaps~\cite{Khachatryan:2015hwa}.
Matching muons to tracks identified in
the silicon tracker results in a $\pt$ resolution for
muons with $20 <\pt < 100\GeV$ of 1.3--2.0\% in the barrel and better than 6\%
in the endcaps. The \pt resolution in the barrel is better than 10\% for muons
with \pt up to 1\TeV~\cite{Chatrchyan:2012xi}.

\section{Signal and background simulation}
\label{sec:mc}

Signal events are generated with
\POWHEG~2.0~\cite{Alioli:2008gx,Nason:2004rx,Frixione:2007vw,Alioli:2010xd,Melia:2011tj} at
next-to-leading order (NLO) in QCD for quark-antiquark processes and
leading order (LO) for quark-gluon processes. This includes
$\ZZ$, $\cPZ\gamma^\ast$, $\cPZ$, and $\gamma^\ast\gamma^\ast$ production
with a constraint of $m_{\ell\ell'} > 4\GeV$ applied to all pairs
of oppositely charged leptons at the generator level to avoid infrared divergences.
The $\Pg\Pg  \to \ZZ$ process is simulated at LO
with \MCFM~v7.0~\cite{Campbell:2010ff}. These samples are scaled to
correspond to cross sections calculated at NNLO in QCD for
$\cPq\cPaq  \to  \ZZ$~\cite{Cascioli:2014yka} (a scaling $K$~factor of 1.1) and at NLO in QCD for
$\Pg\Pg  \to  \ZZ$~\cite{Caola:2015psa} ($K$~factor of 1.7). The
$\Pg\Pg  \to  \ZZ$ process is calculated to
$\mathcal{O}\left(\alpha_s^3\right)$, where $\alpha_s$ is the strong coupling constant,
while the other contributing processes are calculated to
$\mathcal{O}\left(\alpha_s^2\right)$; this higher-order correction
is included because the effect is known to be large~\cite{Caola:2015psa}.
Electroweak \ZZ\ production in association with two jets is generated
with \textsc{Phantom}~v1.2.8~\cite{Ballestrero:2007xq}.

A sample of Higgs boson events is produced in the gluon-gluon fusion
process at NLO with \POWHEG. The Higgs boson
decay is modeled with
\textsc{jhugen}~3.1.8~\cite{Gao:2010qx,Bolognesi:2012,Anderson:2013afp}.
Its cross section is scaled to the NNLO prediction with a $K$~factor of
1.7~\cite{Caola:2015psa}.

{\sloppypar
Samples for background processes containing four prompt leptons in the
final state, $\ttbar\cPZ$ and $\PW\PW\cPZ$ production, are produced with
\MGvATNLO v2.3.3~\cite{mg_amcnlo}.
The $\cPq\cPaq  \to  \PW\cPZ$ process is
generated with \POWHEG.
\par}

Samples with aTGC contributions included are generated at LO with
\textsc{sherpa}~v2.1.1~\cite{Gleisberg:2008ta}.
Distributions from the \textsc{sherpa} samples are normalized such that the
total yield of the SM sample is the same as that of the \POWHEG sample.

{\tolerance=800
The \PYTHIA~v8.175~\cite{Sjostrand:2006za,Sjostrand:2015,Alioli:2010xd} package
is used for parton showering, hadronization, and
the underlying event simulation, with parameters set by the CUETP8M1
tune \cite{Khachatryan:2015pea}, for all samples except the
samples generated with \textsc{sherpa}, which performs these functions
itself. The NNPDF 3.0~\cite{nnpdf}
set is used as the default set of parton distribution functions (PDFs).
For all simulated event samples, the PDFs are calculated to the same
order in QCD as the process in the sample.
\par}

The detector response is simulated using a detailed
description of the CMS detector implemented with the \GEANTfour
package~\cite{GEANT}. The event reconstruction is performed with
the same algorithms used for data.
The simulated samples include additional interactions per bunch crossing,
referred to as pileup.
The simulated events are weighted so that the pileup distribution matches
the data, with an average of about 27 interactions per bunch
crossing.

\section{Event reconstruction}
\label{sec:eventreconstruction}

All long-lived particles---electrons, muons, photons, and charged and neutral
hadrons---in each collision event are identified and reconstructed
with the CMS particle-flow (PF) algorithm~\cite{Sirunyan:2017ulk}
from a combination of the signals from all subdetectors.
Reconstructed electrons~\cite{Khachatryan:2015hwa} and muons~\cite{Chatrchyan:2012xi}
are considered candidates for inclusion in four-lepton final states
if they have $\pt^\Pe > 7\GeV$ and $\abs{\eta^\Pe} < 2.5$ or
$\pt^\Pgm > 5\GeV$ and $\abs{\eta^\Pgm} < 2.4$.

{\tolerance=800
Lepton candidates are also required to originate from the event vertex, defined as the
reconstructed proton-proton interaction vertex with the largest value of
summed physics object $\pt^2$. The physics objects used in the event vertex definition
are the objects returned by a jet finding
algorithm~\cite{Cacciari:2008gp,Cacciari:2011ma} applied to all charged tracks
associated with the vertex, plus the corresponding associated missing transverse
momentum~\cite{CMS-TDR-15-02}. The distance of closest approach between
each lepton track and the event vertex
is required to be less than 0.5\unit{cm} in the plane transverse to the beam axis,
and less than 1\unit{cm} in the direction along the beam axis.
Furthermore, the significance of the three-dimensional impact parameter relative
to the event vertex, $\mathrm{SIP_{3D}}$, is required to satisfy
$\mathrm{SIP_{3D}} \equiv \abs{ \mathrm{IP} / \sigma_\mathrm{IP}} < 10$
for each lepton, where $\mathrm{IP}$ is the distance
of closest approach of each lepton track to the event vertex
and $\sigma_\mathrm{IP}$ is its associated uncertainty.
\par}

Lepton candidates are required to be isolated from other particles in the event. The
relative isolation is defined as
\ifthenelse{\boolean{cms@external}}{
\begin{multline}
        R_\text{iso} = \bigg[ \sum_{\substack{\text{charged} \\ \text{hadrons}}} \!\! \pt \, + \\
                             \max\bigg(0, \sum_{\substack{\text{neutral} \\ \text{hadrons}}} \!\! \pt
                                       + \, \sum_{\text{photons}} \!\! \pt \, - \, \pt^\mathrm{PU}
                                       \bigg)\bigg] \bigg/ \pt^{\ell},
        \label{eq:iso}
\end{multline}
}{
\begin{equation}
        R_\text{iso} = \bigg[ \sum_{\substack{\text{charged} \\ \text{hadrons}}} \!\! \pt \, + \,
                             \max\big(0, \sum_{\substack{\text{neutral} \\ \text{hadrons}}} \!\! \pt
                                       \, + \, \sum_{\text{photons}} \!\! \pt \, - \, \pt^\mathrm{PU}
                                       \big)\bigg] \bigg/ \pt^{\ell},
        \label{eq:iso}
\end{equation}
}
where the sums run over the charged and neutral hadrons and photons identified
by the PF algorithm, in a cone defined by
$\Delta R \equiv \sqrt{\smash[b]{\left(\Delta\eta\right)^2 + \left(\Delta\phi\right)^2}} < 0.3$
around the lepton trajectory. Here $\phi$ is the azimuthal angle in radians.
To minimize the contribution of charged particles from pileup to the isolation calculation,
charged hadrons are included only if they originate from the
event vertex. The contribution of
neutral particles from pileup is $\pt^\mathrm{PU}$. For electrons, $\pt^\mathrm{PU}$
is evaluated with the ``jet area'' method described in Ref.~\cite{Cacciari:2007fd};
for muons, it is taken
to be half the sum of the $\pt$ of all charged particles in the cone originating
from pileup vertices. The factor
one-half accounts for the expected ratio of charged to neutral particle energy
in hadronic interactions. A lepton is considered isolated if
$R_\text{iso} < 0.35$.

The lepton reconstruction, identification, and isolation efficiencies
are measured with a ``tag-and-probe''
technique~\cite{CMS:2011aa} applied to a sample of $\cPZ  \to  \ell^+\ell^-$ data events.
The measurements are performed in several bins of $\pt^{\ell} $ and $ |\eta^\ell|$.
The electron reconstruction and selection efficiency in the ECAL barrel (endcaps) varies from
about 85\% (77\%) at $\PT^{\Pe} \approx 10\GeV$
to about 95\% (89\%) for $\PT^{\Pe} \geq 20\GeV$,
while in the barrel-endcap transition region this efficiency is about 85\% averaged
over all electrons with $\pt^{\Pe} > 7\GeV$.
The muons are reconstructed and identified with efficiencies above ${\sim}98\%$
within $\abs{\eta^{\Pgm}} < 2.4$.

\section{Event selection}

The primary triggers for this analysis require the presence of a
pair of loosely isolated leptons of the same or different
flavors~\cite{Khachatryan:2016bia}.
The highest \pt lepton must have $\pt^\ell > 17\GeV$, and the
subleading lepton must have
$\pt^\Pe > 12\GeV$ if it is an electron or $\pt^\Pgm > 8\GeV$
if it is a muon. The tracks of the triggering leptons are required to originate within
2~mm of each other in the plane transverse to the beam axis. Triggers
requiring a triplet of lower-\pt leptons
with no isolation criterion, or a single high-\pt electron or muon, are also used.
An event is used if it passes any trigger regardless of the decay channel.
The total trigger efficiency for events within the acceptance of this
analysis is greater than 98\%.

The four-lepton candidate selections are based on those used in
Ref.~\cite{Chatrchyan:2013mxa}. A signal event must contain at least two
$\cPZ/\gamma^{\ast}$ candidates, each formed from an oppositely charged
pair of isolated electron candidates or muon candidates.
Among the four leptons, the highest \pt lepton must have $\pt > 20\GeV$, and
the second-highest \pt lepton must have $\pt^\Pe > 12\GeV$ if it is an electron
or $\pt^\Pgm > 10\GeV$ if it is a muon.
All leptons are required to be separated from each other by
$\Delta R \left(\ell_1, \ell_2 \right) > 0.02$,
and electrons are required to be separated from muons by
$\Delta R \left(\Pe, \mu \right) > 0.05$.

Within each event, all
permutations of leptons giving a valid pair of $\cPZ/\gamma^{\ast}$
candidates are considered separately.
Within each $\elfour$ candidate, the
dilepton candidate with an invariant mass closest to 91.2\GeV, taken as
the nominal $\cPZ$ boson mass~\cite{Olive:2016xmw}, is denoted $\cPZ_1$ and is required to have a
mass greater than 40\GeV. The other dilepton candidate is denoted $\cPZ_2$.
Both $m_{Z_1}$ and $m_{Z_2}$ are required to be less than 120\GeV.
All pairs of oppositely charged leptons in the $4\ell$ candidate are required to
have $m_{\ell \ell'} > 4\GeV$ regardless of their flavor.

If multiple $\elfour$
candidates within an event pass all selections,
the one with $m_{\cPZ_1}$ closest to
the nominal $\cPZ$ boson mass is chosen. In the rare case
of further ambiguity, which may arise in less than 0.5\% of events when
five or more passing lepton candidates are found, the $\cPZ_2$ candidate
that maximizes the scalar $\pt$ sum of the four leptons is chosen.

Additional requirements
are applied to select events for measurements of specific processes.
The $\pp  \to  \cPZ\cPZ$ cross section is measured using events where both
$m_{\cPZ_1}$ and $m_{\cPZ_2}$ are greater than 60\GeV.
The $\cPZ  \to \elfour$
branching fraction is measured using events with
$80 < m_{\elfour} < 100\GeV$, a range chosen
to retain most of the decays in the resonance while removing most other
processes with four-lepton final states. Decays of the $\cPZ$ bosons to
$\PGt$ leptons with subsequent decays to electrons and muons are heavily
suppressed by requirements on lepton $\pt$, and the contribution of such
events is less than 0.5\% of the total $\ZZ$ yield. If these events pass the
selection requirements of the analysis, they are considered signal, while they
are not considered at generator level in the cross section unfolding procedure.
Thus, the correction for possible $\tau$ decays is included in the efficiency
calculation.

\section{Background estimate}

The major background contributions arise from
$\cPZ$ boson and $\PW\cPZ$ diboson production in association with
jets and from \ttbar production.
In all these cases, particles from jet fragmentation satisfy both lepton identification and
isolation criteria, and are thus misidentified as signal leptons.

The probability for such objects to be selected
is measured from a sample of
$\cPZ + \ell_\text{candidate}$ events, where $\cPZ$ denotes a pair of
oppositely charged, same-flavor leptons that pass all analysis requirements and
satisfy $\abs{ m_{\ell^+\ell^-} - m_{\cPZ}} < 10\GeV$, where
$m_\cPZ$ is the nominal $\cPZ$ boson mass. Each event in this sample must have exactly one
additional object $\ell_\text{candidate}$ that passes relaxed identification requirements with
no isolation requirements applied. The misidentification probability for
each lepton flavor, measured in bins of lepton candidate $\pt$ and $\eta$,
is defined as the ratio of the number of candidates that pass the final
isolation and identification requirements to the total number in the sample.
The number of $\cPZ + \ell_\text{candidate}$ events is corrected for the contamination
from $\PW\cPZ$ production and $\ZZ$ production in which one lepton is not
reconstructed. These events have a third genuine, isolated lepton that must be excluded
from the misidentification probability calculation. The WZ contamination is suppressed by requiring the missing
transverse momentum $\ptmiss$ to be below 25\GeV. The $\ptmiss$ is defined
as the magnitude of the missing transverse momentum vector $\ptvecmiss$,
the projection onto the plane transverse to the beams of the negative
vector sum of the momenta of all reconstructed PF candidates in the event,
corrected for the jet energy scale.
Additionally, the transverse mass calculated with $\ptvecmiss$ and the
$\ptvec$ of $\ell_\text{candidate}$,
$m_{\mathrm{T}} \equiv \sqrt{\smash[b]{(\pt^\ell + \ptmiss )^2 -
(\ptvecl + \ptvecmiss )^2}}$, is required to be less than 30\GeV. The residual contribution
of $\PW\cPZ$ and $\cPZ\cPZ$ events, which may be up to a few percent of the
events with $\ell_\text{candidate}$ passing all selection criteria, is estimated
from simulation and subtracted.

To account for all sources of background events, two control samples are
used to estimate the number of background events
in the signal regions. Both are defined to contain events with
a dilepton candidate satisfying all requirements ($\cPZ_1$) and
two additional lepton candidates $\ell^{\prime +}\ell^{\prime -}$.
In one control sample, enriched in $\PW\cPZ$ events, one
$\ell^{\prime}$ candidate is required to satisfy the full
identification and isolation criteria and the other must fail the full criteria
and instead satisfy only the relaxed ones; in the other, enriched in
$\cPZ$+jets events, both $\ell^{\prime}$
candidates must satisfy the relaxed criteria, but fail the full criteria.
The additional leptons must have opposite charge and the same
flavor ($\Pe^{\pm}\Pe^{\mp}, \Pgm^{\pm}\Pgm^{\mp}$).
From this set of events, the expected number of background events in the
signal region, denoted ``$\cPZ + \text{X}$'' in the figures, is obtained
by scaling the number of observed $\cPZ_1+\ell^{\prime +}\ell^{\prime -}$ events
by the misidentification probability for each lepton failing the selection.
It is found to be approximately 4\% of the total expected yield.
The procedure is described in more detail in Ref.~\cite{Chatrchyan:2013mxa}.

In addition to these nonprompt backgrounds,
$\ttbar\cPZ$ and $\PW\PW\cPZ$ processes contribute a smaller number of events
with four prompt leptons, which is estimated from simulated samples
to be around 1\% of the expected $\ZZ \to 4\ell$ yield.
In the $\cPZ \to 4\ell$ selection, the contribution from these backgrounds
is negligible.
The total background contributions to the $\cPZ \to \elfour$ and
$\cPZ\cPZ \to \elfour$ signal regions are summarized in
Section~\ref{sec:xsec}.

\section{Systematic uncertainties}\label{sec:systematics}

The major sources of systematic uncertainty and their effect on the
measured cross sections are summarized in Table~\ref{table:systematics}.
In both data and simulated event samples, trigger efficiencies are evaluated with
a tag-and-probe technique. The ratio of data to simulation is applied
to simulated events, and the size of the resulting change in expected yield is
taken as the uncertainty in the determination of the trigger efficiency.
This uncertainty is
around 2\% of the final estimated yield. For
$\cPZ  \to  4\Pe$ events, the uncertainty increases to 4\%.

\begin{table}[htb]
\centering
\topcaption{
  The contributions of each source of systematic uncertainty in the
  cross section measurements. The integrated luminosity uncertainty, and the PDF and scale
  uncertainties, are considered separately. All other uncertainties are added
  in quadrature into a single systematic uncertainty. Uncertainties that vary by
  decay channel are listed as a range.
}
\begin{tabular}{lcc}
\hline
Uncertainty & $\cPZ  \to  4\ell$ & $\cPZ\cPZ  \to  4\ell$  \\
\hline
Lepton efficiency         & 6--10\%      & 2--6\%     \\
Trigger efficiency        & 2--4\%       & 2\%        \\
Statistical (simulation)  & 1--2\%       & 0.5\%      \\
Background                & 0.6--1.3\%   & 0.5--1\%   \\
Pileup                    & 1--2\%       & 1\%        \\[\cmsTabSkip]
PDF                       & 1\%          & 1\%        \\
$\mu_\mathrm{R}$, $\mu_\mathrm{F}$          & 1\%          & 1\%        \\[\cmsTabSkip]
Integrated luminosity     & 2.5\%        & 2.5\%      \\
\hline
\end{tabular}
\label{table:systematics}
\end{table}

The lepton identification, isolation, and track reconstruction efficiencies in simulation are corrected
with scaling factors derived with a tag-and-probe method and applied as a
function of lepton $\pt$ and $\eta$.
To estimate the uncertainties associated with the tag-and-probe technique, the total yield is
recomputed with the scaling factors varied up and down by the tag-and-probe
fit uncertainties. The uncertainties associated with lepton efficiency in
the $\ZZ  \to  \elfour$ ($\cPZ  \to  \elfour$)
signal regions are found to be 6(10)\% in the $4\Pe$, 3(6)\%
in the  $2\Pe 2\mu$, and 2(7)\% in the $4\mu$ final states.
These uncertainties are
higher for $\cPZ  \to  \elfour$ events because the leptons
generally have lower \pt, and the samples used in the tag-and-probe method
have fewer events and more contamination from nonprompt leptons in this
low-\pt region.

Uncertainties due to the effect of factorization ($\mu_\mathrm{F}$) and
renormalization ($\mu_\mathrm{R}$) scale choices on the $\ZZ \to \elfour$
acceptance are evaluated with \POWHEG and \MCFM by varying the scales up
and down by a factor of two with respect to the default values
$\mu_\mathrm{F} = \mu_\mathrm{R} = m_{\cPZ\cPZ}$. All combinations are considered
except those in which $\mu_\mathrm{F}$ and $\mu_\mathrm{R}$ differ by a factor of
four. Parametric uncertainties
(PDF$+ \alpha_s$) are evaluated according to the \textsc{pdf4lhc}
prescription~\cite{Butterworth:2015oua} in the acceptance calculation,
and with NNPDF3.0 in the cross section calculations.
An additional theoretical uncertainty arises from scaling the \POWHEG
$\cPq\cPaq  \to  \ZZ$ simulated sample from its NLO cross section to the
NNLO prediction, and the \MCFM $\Pg\Pg  \to  \ZZ$ samples
from their LO cross sections to the NLO predictions. The change in the
acceptance corresponding to this scaling procedure is found to be 1.1\%.
All these theoretical uncertainties are added in quadrature.

The largest uncertainty in the estimated background yield arises from
differences in sample composition between the $\cPZ + \ell_\text{candidate}$ control sample
used to calculate the lepton misidentification probability and the
$\cPZ + \ell^+\ell^-$ control sample. A further uncertainty arises
from the limited number of events in the $\cPZ + \ell_\text{candidate}$ sample. A
systematic uncertainty of 40\% is applied to the lepton misidentification probability
to cover both effects. The size of this uncertainty varies by channel, but
is less than 1\% of the total expected yield.

The uncertainty in the integrated luminosity of the data sample
is 2.5\%~\cite{CMS-PAS-LUM-17-001}.

\section{Cross section measurements}
\label{sec:xsec}

\begin{figure}[htbp]
\centering
\includegraphics[width=0.48\textwidth]{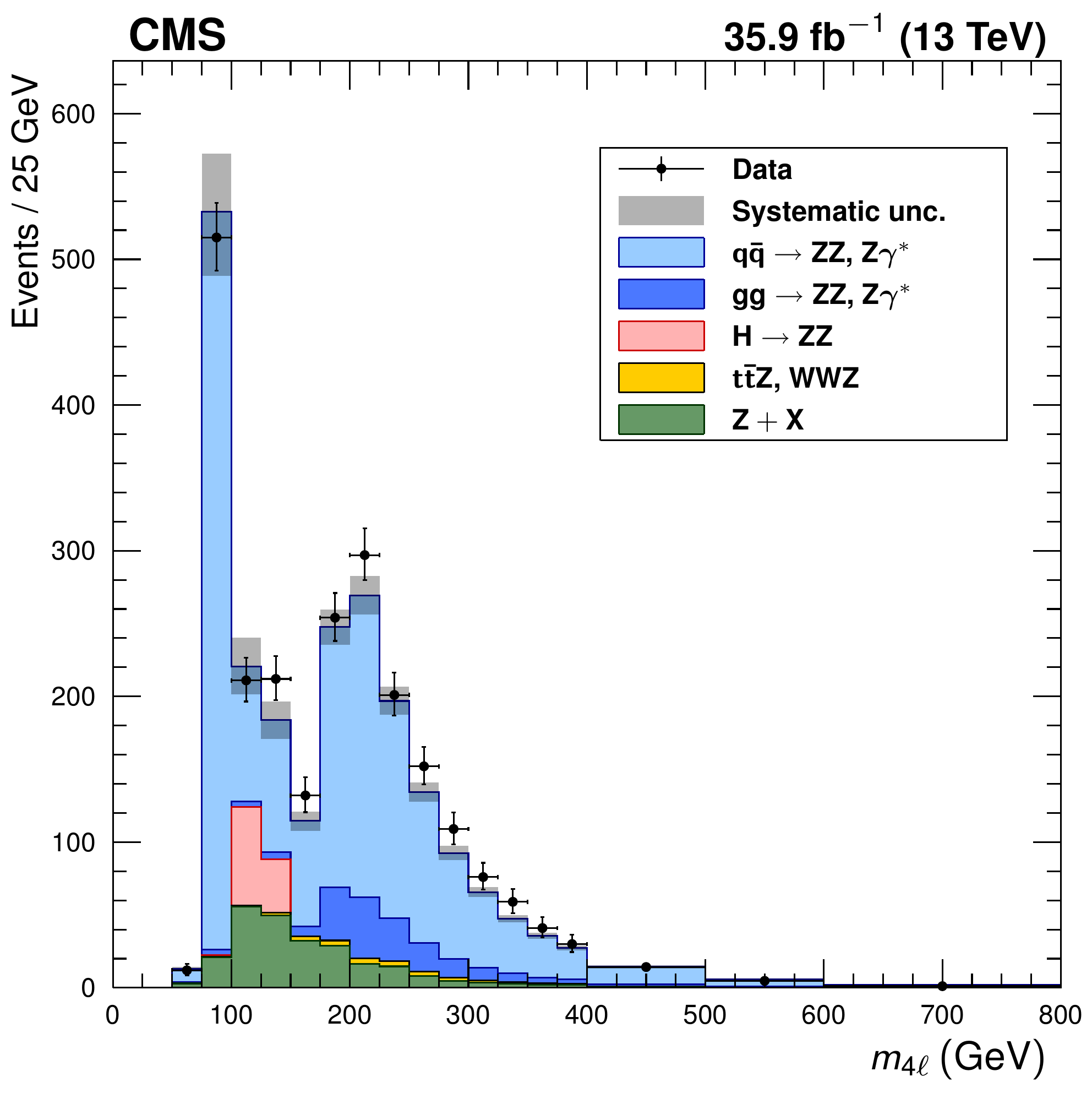}
\includegraphics[width=0.48\textwidth]{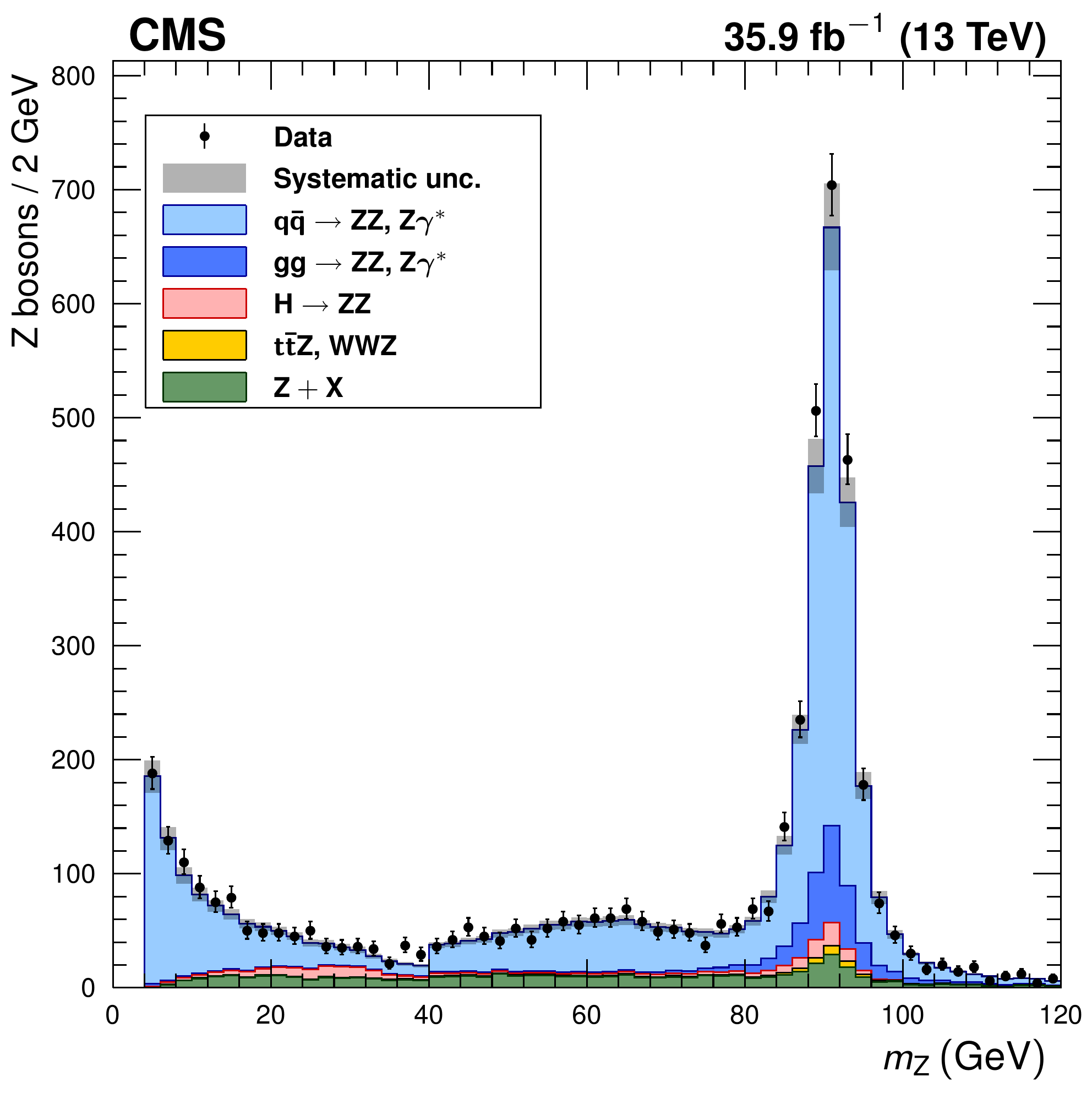}
\caption{
Distributions of (\cmsLeft) the four-lepton invariant mass $m_{\elfour}$
and (\cmsRight) the dilepton invariant mass of all
$\cPZ/\gamma^\ast$ bosons in selected
four-lepton events. Both selected dilepton candidates are included in each event.
In the $m_{\elfour}$ distribution, bin contents
are normalized to a bin width of 25\GeV; horizontal
bars on the data points show the range of the corresponding bin.
Points represent the data, while filled histograms represent
the SM prediction and background estimate. Vertical bars on the data points
show their statistical uncertainty. Shaded grey regions around the predicted yield
represent combined statistical, systematic, theoretical, and integrated
luminosity uncertainties.
}
\label{fig:results_full}
\end{figure}

\begin{figure}[htbp]
\centering
\includegraphics[width=0.48\textwidth]{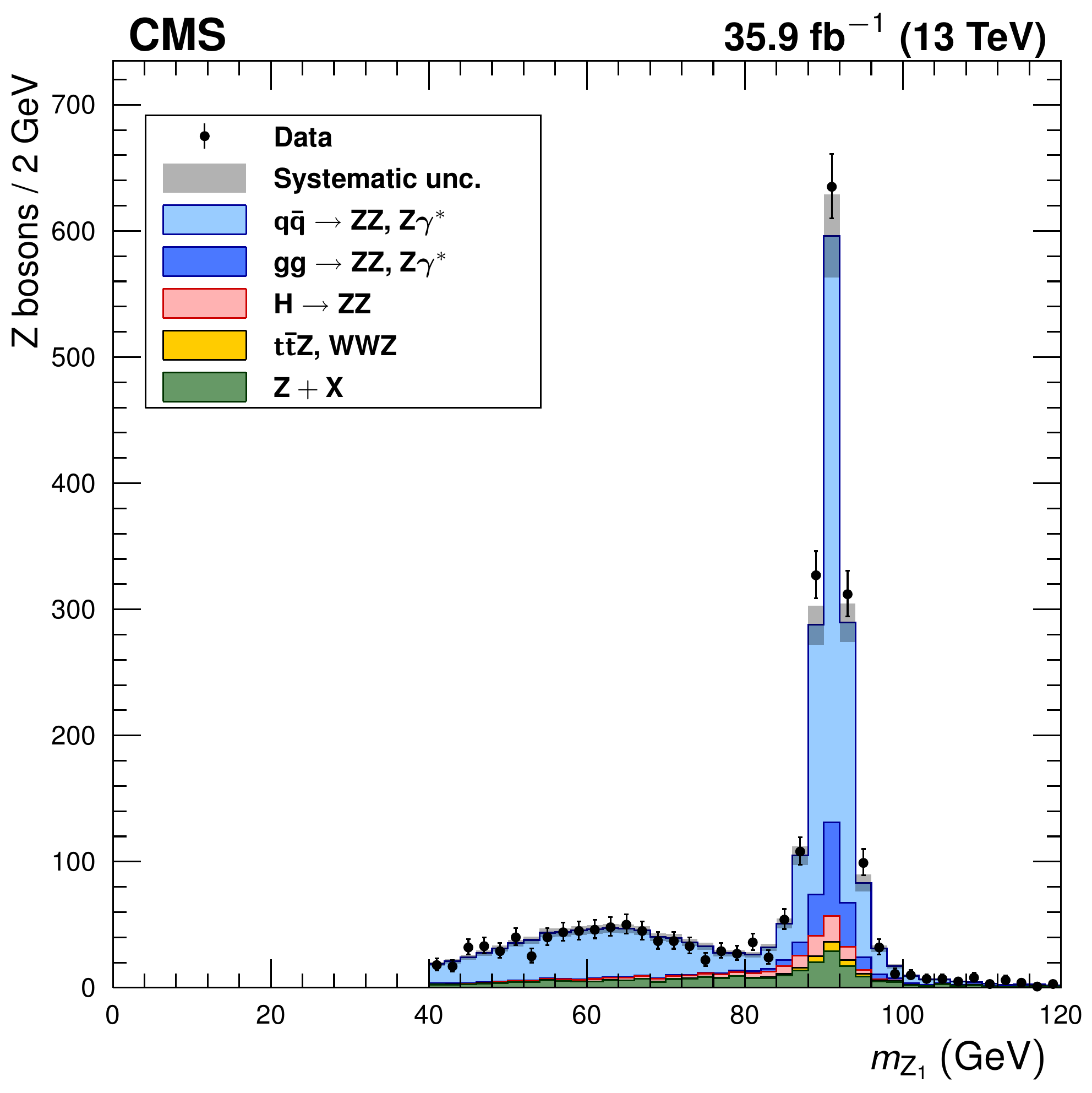}
\includegraphics[width=0.48\textwidth]{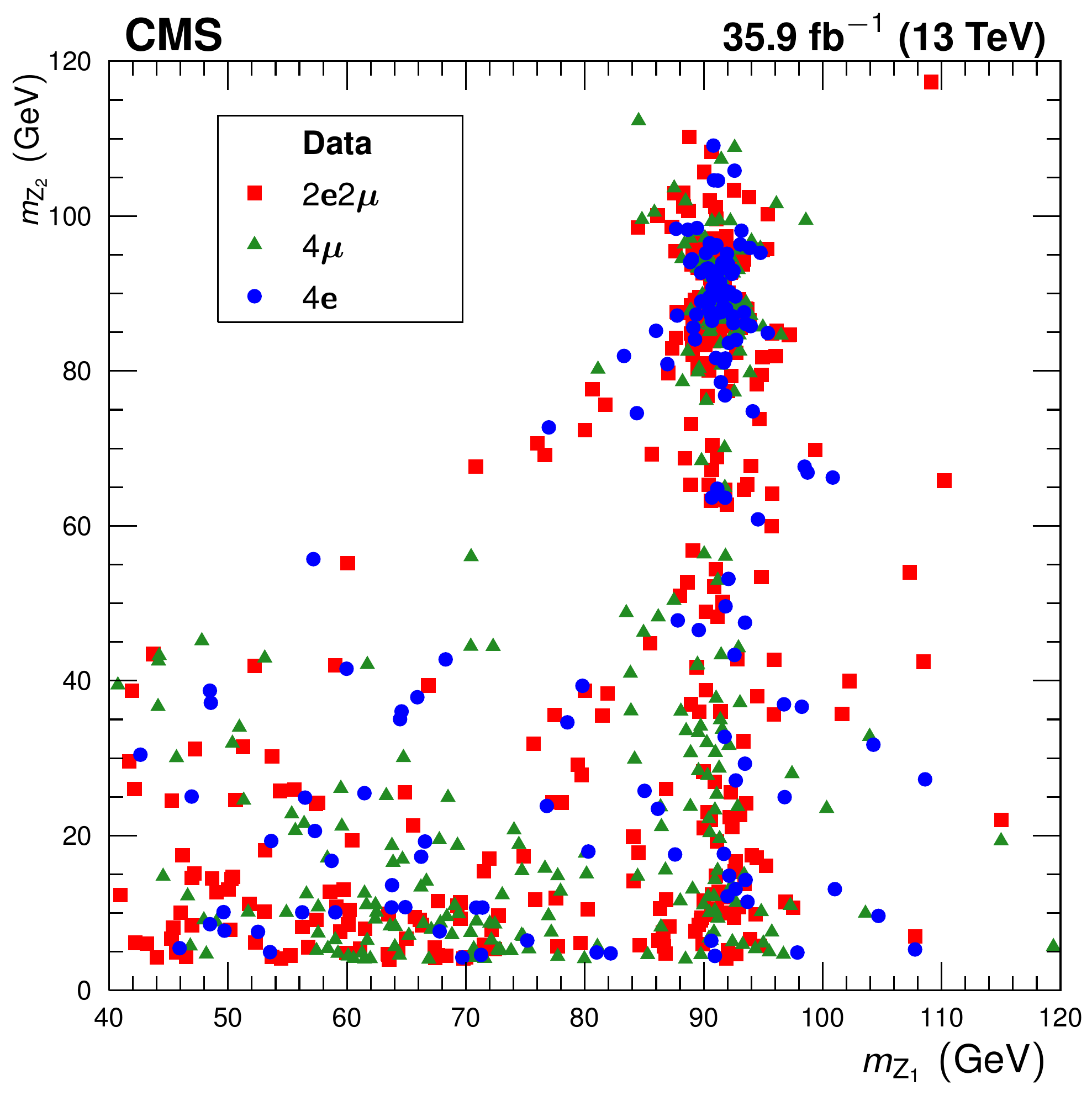}
\caption{
(\CMSLeft): the distribution of the reconstructed mass of $\cPZ_1$, the
dilepton candidate closer to the nominal {\cPZ} boson mass.
Points represent the data, while filled histograms represent
the SM prediction and background estimate. Vertical bars on the data points
show their statistical uncertainty. Shaded grey regions around the predicted yield
represent combined statistical, systematic, theoretical, and integrated
luminosity uncertainties.
(\CMSRight): the reconstructed $m_{\cPZ_2}$ plotted
against the reconstructed $m_{\cPZ_1}$ in data events, with
distinctive markers for each final state. For readability, only every fourth
event is plotted.
}
\label{fig:results_full_Z}
\end{figure}

The distributions of the four-lepton mass and the masses of
the $\cPZ_1$ and $\cPZ_2$ candidates are shown in Fig.~\ref{fig:results_full}.
The expected distributions describe the data well within uncertainties.
The SM predictions include nonresonant $\cPZ\cPZ$ predictions,
production of the SM Higgs
boson with mass 125\GeV~\cite{Aad:2015zhl}, and
resonant $\cPZ \to \elfour$ production.
The backgrounds estimated from
data and simulation are also shown.
The reconstructed invariant mass of the $\cPZ_1$ candidates, and a scatter plot
showing the correlation between $m_{\cPZ_2}$ and $m_{\cPZ_1}$ in data events,
are shown in Fig.~\ref{fig:results_full_Z}. In the scatter plot,
clusters of events corresponding to $\cPZ\cPZ \to \elfour$,
$\cPZ\gamma^\ast \to \elfour$, and $\cPZ \to \elfour$ production can be seen.

\begin{figure}[htbp]
\centering
\includegraphics[width=0.48\textwidth]{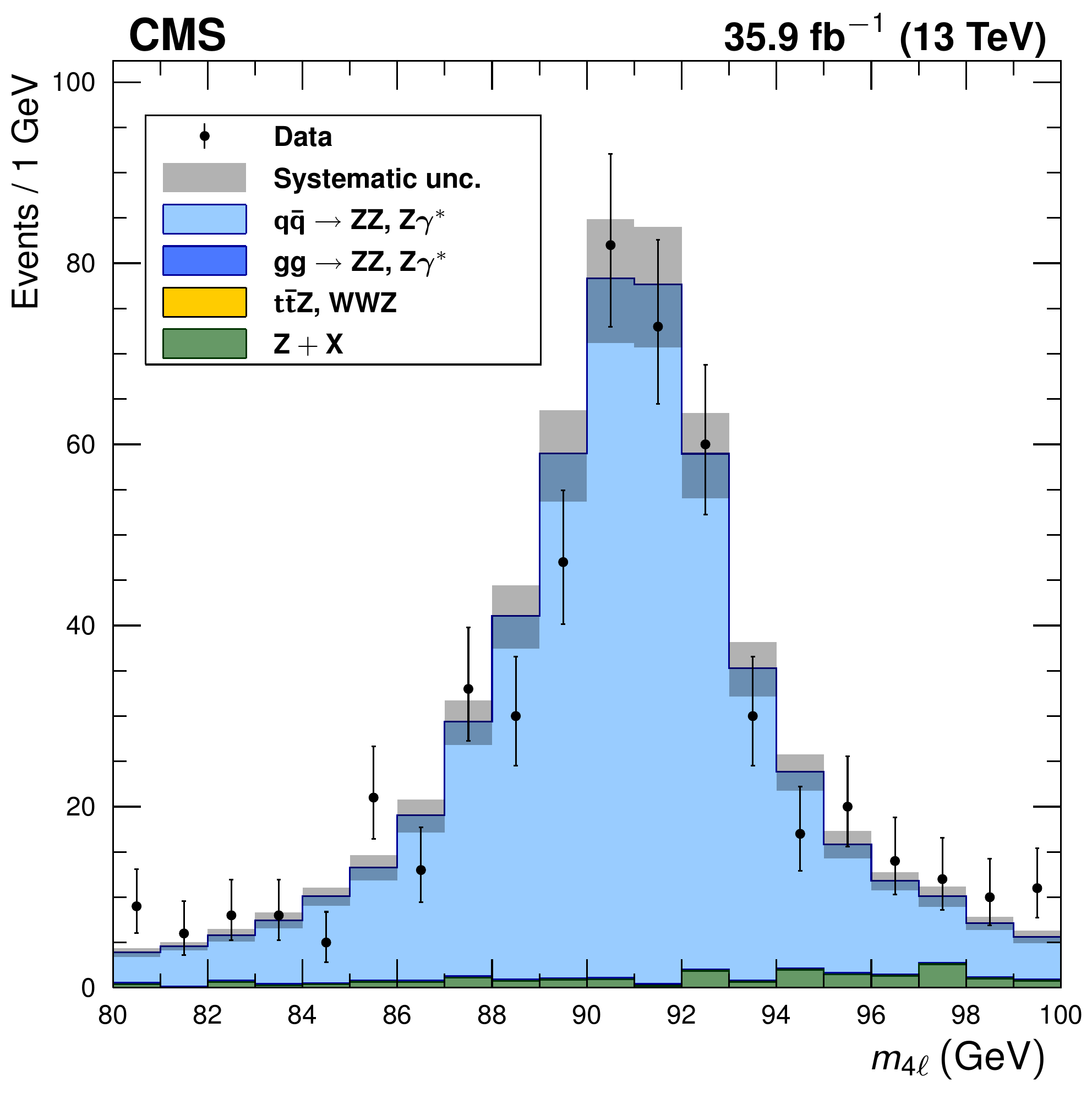}
\includegraphics[width=0.48\textwidth]{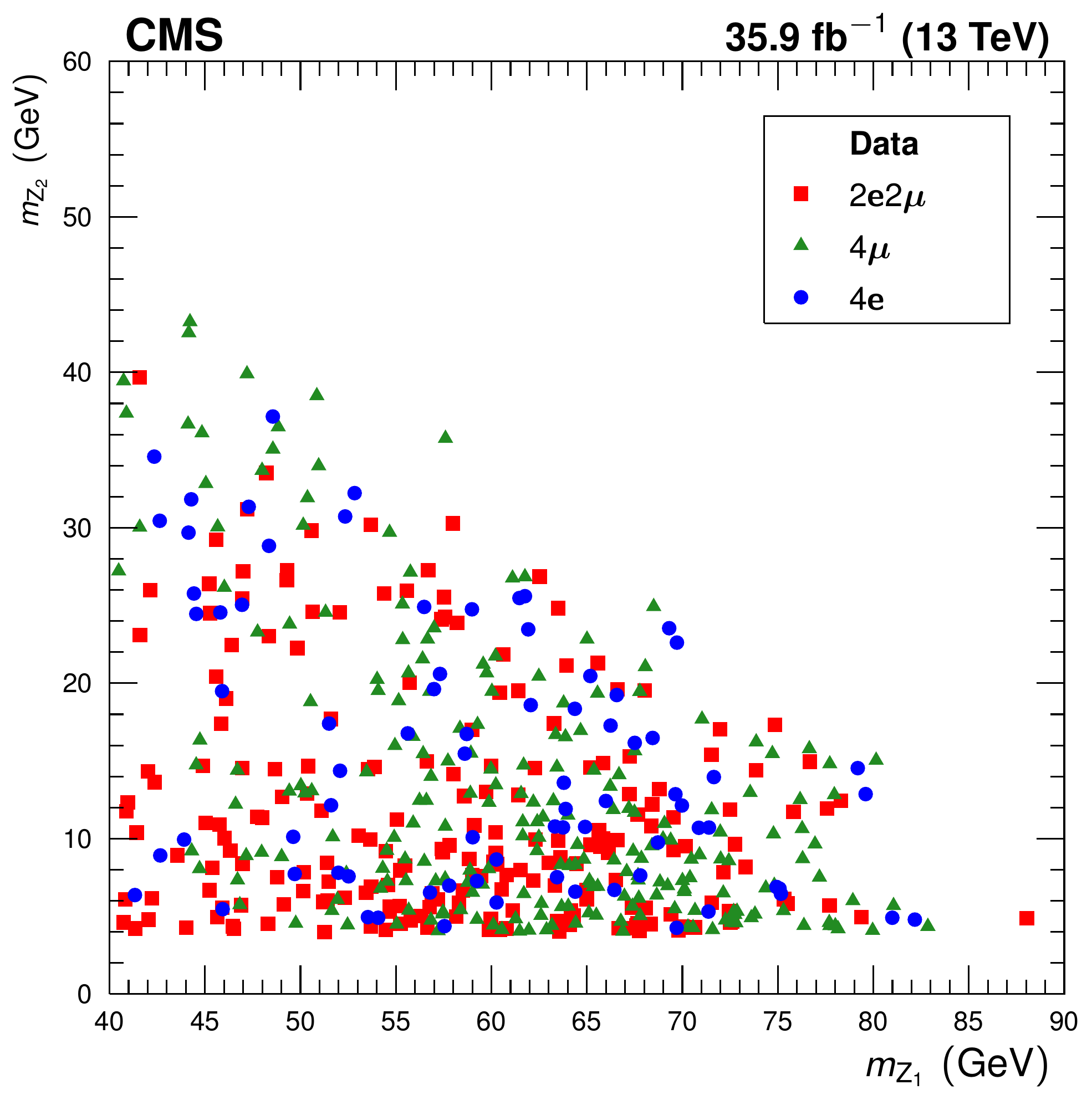}
\caption{
(\CMSLeft): the distribution of the reconstructed four-lepton mass
$m_{\elfour}$ for events selected with $80 < m_{\elfour} < 100\GeV$.
Points represent the data, while filled histograms represent
the SM prediction and background estimate. Vertical bars on the data points
show their statistical uncertainty. Shaded grey regions around the predicted yield
represent combined statistical, systematic, theoretical, and integrated
luminosity uncertainties.
(\CMSRight): the reconstructed $m_{\cPZ_2}$ plotted
against the reconstructed $m_{\cPZ_1}$ for all data events selected with
$m_{\elfour}$ between 80 and 100\GeV, with
distinctive markers for each final state. }
\label{fig:results_z4l}
\end{figure}

{\tolerance=1200
The four-lepton invariant mass distribution below 100\GeV
is shown in Fig.~\ref{fig:results_z4l} (\cmsLeft).
Figure~\ref{fig:results_z4l} (\cmsRight) shows $m_{\cPZ_2}$ plotted against
$m_{\cPZ_1}$ for events with
$m_{\elfour}$ between 80 and 100\GeV,
and the observed and expected event yields
in this mass region are given in Table~\ref{table:results_z4l}.
The yield of events in the $4\Pe$ final state is significantly lower than in the $4\Pgm$
final state because minimum $\pt$ thresholds are higher for electrons than for muons,
and inefficiencies in the detection of low-$\pt$ leptons affect
electrons more strongly than they affect muons.
\par}

\begin{table*}[htbp]
\centering
\topcaption{ The observed and expected yields of four-lepton
 events in the mass region
$80 < m_{\elfour} < 100\GeV$
and estimated yields of background
events, shown for each final state and
summed in the total expected yield.
The first uncertainty is statistical, the second one is systematic.
The systematic uncertainties do not include the uncertainty in the integrated
luminosity.
}
\begin{tabular}{lcccr}
        \hline
Final & Expected &  Background   & Total & Observed \\
state & $N_{\elfour}$ &  & expected  & \\[0.2ex]
        \hline
        $4\Pgm$       & $ 224 \pm 1 \pm 16 $ & $ 7 \pm 1 \pm 2 $ & $ 231 \pm 2 \pm 17 $ & $ 225 $ \\
        $2\Pe 2\Pgm$  & $ 207 \pm 1 \pm 14 $ & $ 9 \pm 1 \pm 2 $ & $ 216 \pm 2 \pm 14 $ & $ 206 $ \\
        $4\Pe$        & $ 68 \pm 1 \pm 8 $ & $ 4 \pm 1 \pm 2 $ & $ 72 \pm 1 \pm 8 $ & $ 78 $ \\[\cmsTabSkip]

        Total         & $ 499  \pm 2  \pm 32 $  & $ 19  \pm 2  \pm 5 $  & $ 518  \pm 3  \pm 33 $  & $ 509 $  \\
        \hline
      \end{tabular}
  \label{table:results_z4l}
\end{table*}

\begin{figure}[htbp]
\centering
\includegraphics[width=0.48\textwidth]{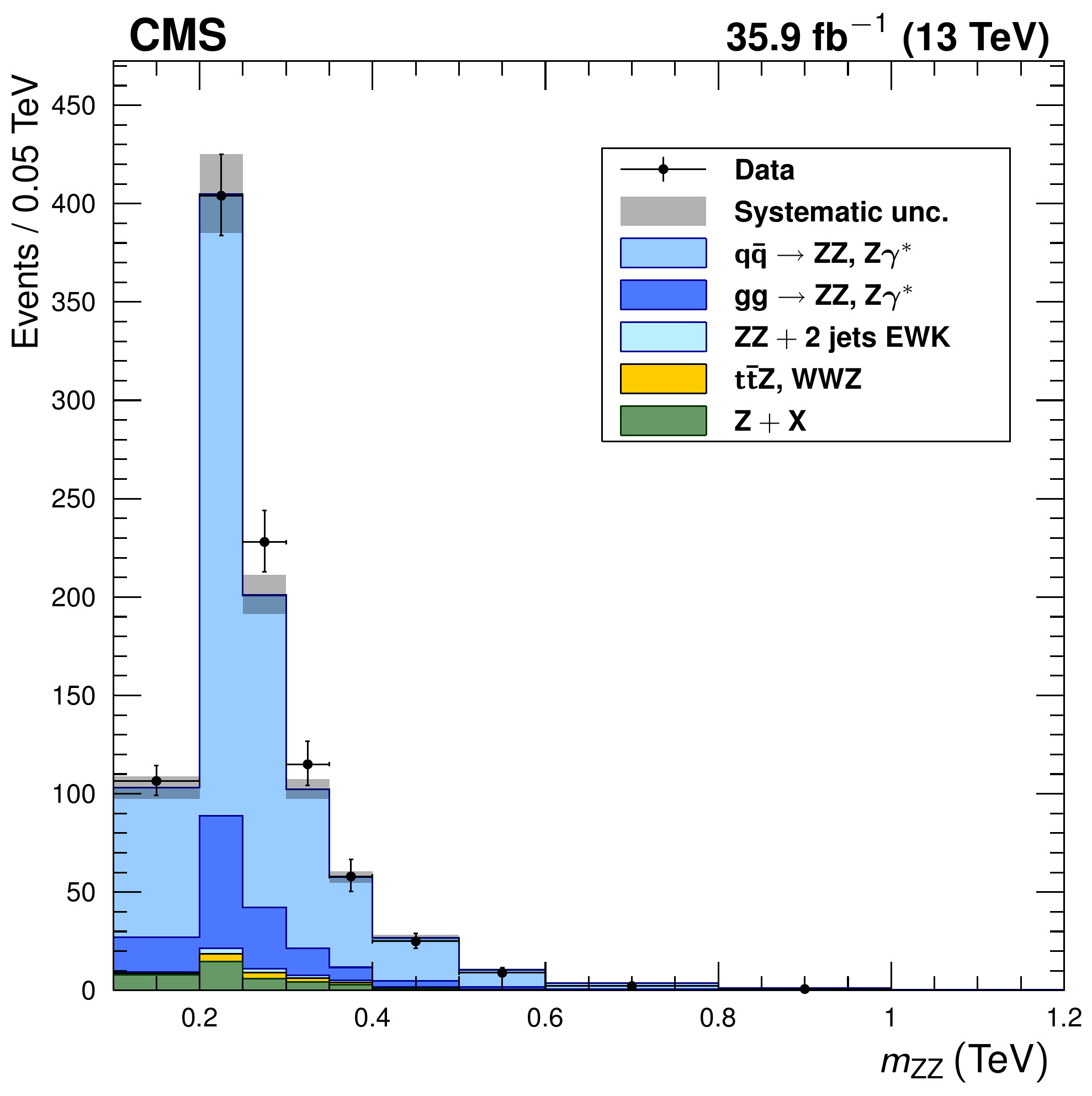}
\includegraphics[width=0.48\textwidth]{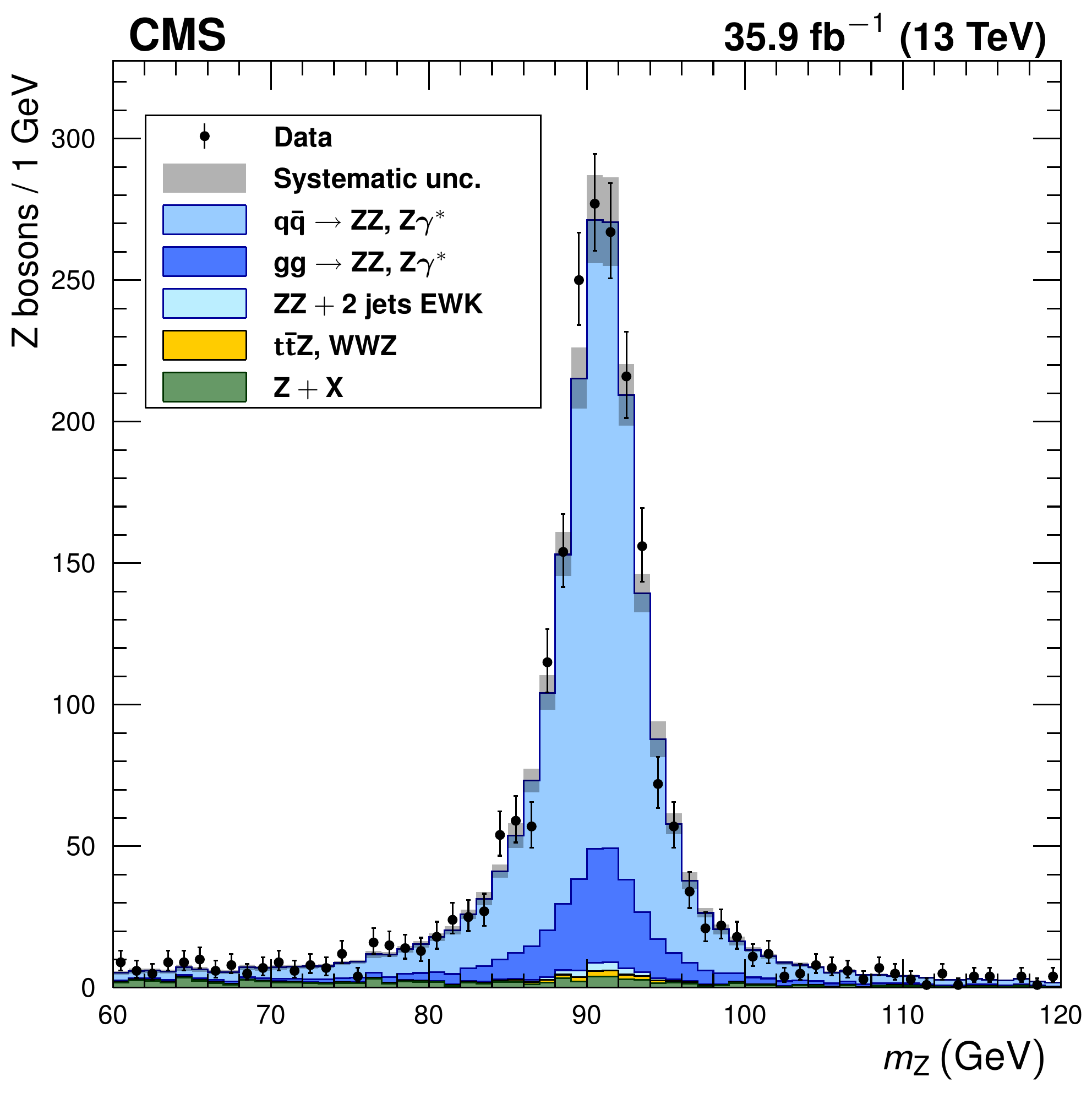}
\caption{
Distributions of (\cmsLeft) the four-lepton invariant mass $m_{\ZZ}$
and (\cmsRight) dilepton candidate mass for four-lepton events selected with
both $\cPZ$ bosons on-shell.
Points represent the data, while filled histograms represent
the SM prediction and background estimate. Vertical bars on the data points
show their statistical uncertainty. Shaded grey regions around the predicted yield
represent combined statistical, systematic, theoretical, and integrated
luminosity uncertainties.
In the $m_{\ZZ}$ distribution, bin contents
are normalized to the bin widths, using a unit bin size of 50\GeV; horizontal
bars on the data points show the range of the corresponding bin.
}
\label{fig:results_smp}
\end{figure}

The reconstructed four-lepton invariant mass is shown  in Fig.~\ref{fig:results_smp} (\cmsLeft)
for events with two on-shell $\cPZ$ bosons.
Figure~\ref{fig:results_smp} (\cmsRight) shows the invariant mass distribution for all
$\cPZ$~boson candidates in these events.  The corresponding observed and expected
yields are given in Table~\ref{table:results_smp}.

\begin{table*}[htbp]
\centering
  \topcaption{ The observed and expected yields of $\cPZ\cPZ$ events,
    and estimated yields of background
    events, shown for each final state and
    summed in the total expected yield. The first uncertainty
    is statistical, the second one is systematic. The systematic uncertainties
    do not include the uncertainty in the integrated luminosity.
    }
  \begin{tabular}{lcccr}
    \hline
    Decay   & Expected  &  Background   & Total & Observed \\
    channel & $N_{\elfour}$ &  & expected  & \\[0.2ex]
    \hline

        $4\Pgm$       & $ 301 \pm 2 \pm 9 $ & $ 10 \pm 1 \pm 2 $ & $ 311 \pm 2 \pm 9 $ & $ 335 $ \\
        $2\Pe 2\Pgm$  & $ 503 \pm 2 \pm 19 $ & $ 31 \pm 2 \pm 4 $ & $ 534 \pm 3 \pm 20 $ & $ 543 $ \\
        $4\Pe$        & $ 205 \pm 1 \pm 12 $ & $ 20 \pm 2 \pm 2 $ & $ 225 \pm 2 \pm 13 $ & $ 220 $ \\[\cmsTabSkip]
        Total         & $ 1009  \pm 3  \pm 36 $  & $ 60  \pm 3  \pm 8 $  & $ 1070  \pm 4  \pm 37 $  & $ 1098 $  \\
        \hline
  \end{tabular}
  \label{table:results_smp}
\end{table*}

The observed yields are used to evaluate the
$\pp  \to  \cPZ  \to  \elfour$
and $\pp  \to  \ZZ  \to \elfour$
production cross sections
from a combined fit to the number of observed
events in all the final states. The likelihood is a combination of
individual channel likelihoods for the signal and background hypotheses
with the statistical and systematic uncertainties in the form of scaling
nuisance parameters.
The fiducial cross section is measured by scaling the cross section in the
simulation by the ratio of the measured and predicted event yields given
by the fit.

The definitions for the fiducial phase spaces for the
$\cPZ \to \elfour$ and $\ZZ \to \elfour$
cross section measurements are given in Table~\ref{table:fiducial_cuts}.
In the $\ZZ \to \elfour$ case, the $\cPZ$ bosons used in the fiducial definition
are built by pairing final-state leptons using the same algorithm as is used to
build $\cPZ$ boson candidates from reconstructed leptons.
The generator-level leptons used for the fiducial cross section calculation
are ``dressed'' by adding the momenta of generator-level photons within
$\Delta R\left(\ell,\gamma\right) < 0.1$ to their momenta.

\begin{table*}[htbp]
\centering
\topcaption{
Fiducial definitions for the reported cross sections.
The common requirements are applied for both measurements.
}
\begin{tabular}{ll}
\hline
Cross section measurement & Fiducial requirements \\
\hline
Common requirements & $\pt^{\ell_1} > 20\GeV$,
  $\pt^{\ell_2} > 10\GeV$,
  $\pt^{\ell_{3,4}} > 5\GeV$, \\
 & \\ [-1.9ex]
 & $\abs{\eta^{\ell}} < 2.5$,
  $m_{\ell\ell} > 4\GeV$ (any opposite-sign same-flavor pair) \\[\cmsTabSkip]
$\cPZ \to \elfour$ & $m_{\cPZ_1} > 40\GeV$ \\
 & $80 < m_{\elfour} < 100\GeV$ \\[\cmsTabSkip]
$\cPZ\cPZ \to \elfour$   & $60 < \left(m_{\cPZ_1}, m_{\cPZ_2}\right) < 120\GeV$ \\
\hline
\end{tabular}
\label{table:fiducial_cuts}
\end{table*}

The measured cross sections are
\ifthenelse{\boolean{cms@external}}{
\begin{equation}
\label{xs:z4l}
\begin{aligned}
  \sigma_{\text{fid}}&(\pp \to \cPZ \to \elfour) \\&= 31.2 _{-1.4}^{+1.5} \stat _{-1.9}^{+2.1} \syst \pm 0.8 \lum \unit{fb},\\
  \sigma_{\text{fid}}&(\pp \to \ZZ \to \elfour) \\&= 40.9 \pm 1.3 \stat \pm 1.4 \syst \pm 1.0 \lum \unit{fb}.
  \end{aligned}
\end{equation}
}{
\begin{equation}
\label{xs:z4l}
\begin{aligned}
  \sigma_{\text{fid}} (\pp \to \cPZ \to \elfour) &= 31.2 _{-1.4}^{+1.5} \stat _{-1.9}^{+2.1} \syst \pm 0.8 \lum \unit{fb},\\
  \sigma_{\text{fid}} (\pp \to \ZZ \to \elfour) &= 40.9 \pm 1.3 \stat \pm 1.4 \syst \pm 1.0 \lum \unit{fb}.
  \end{aligned}
\end{equation}
}
The $\pp  \to  \cPZ  \to  \elfour$
fiducial cross section can be compared to
$27.9^{+1.0}_{-1.5} \pm 0.6\unit{fb}$ calculated at NLO in QCD with \POWHEG using
the same settings as used for the simulated sample described in Section~\ref{sec:mc}, with
dynamic scales $\mu_\mathrm{F} = \mu_\mathrm{R} = m_{\elfour}$.
The uncertainties correspond to scale and PDF variations, respectively.
The $\ZZ$ fiducial cross section can be compared to
$34.4^{+0.7}_{-0.6} \pm 0.5\unit{fb}$ calculated with \POWHEG and \MCFM
using the same settings as the simulated samples, or to $36.0_{-0.8}^{+0.9}$
computed with \textsc{matrix} at NNLO. The \POWHEG and \textsc{matrix}
calculations used dynamic scales
$\mu_\mathrm{F} = \mu_\mathrm{R} = m_{\elfour}$, while the
contribution from \MCFM was computed with dynamic scales
$\mu_\mathrm{F} = \mu_\mathrm{R} = 0.5 m_{\elfour}$.

The $\pp  \to  \cPZ  \to  \elfour$ fiducial cross section
is scaled to $\sigma (\pp \to \cPZ) \mathcal{B} (\cPZ \to 4\ell)$ using
the acceptance correction factor
$\mathcal{A} = 0.125 \pm 0.002$, estimated with \POWHEG.
This factor corrects the fiducial $\cPZ  \to  \elfour$ cross section to the
phase space with only the 80--100\GeV mass window and $m_{\elpelm} > 4\GeV$
requirements, and also includes a correction, $0.96 \pm 0.01$,
for the contribution of nonresonant four-lepton production to
the signal region. The uncertainty takes into account the interference between
doubly- and singly-resonant diagrams. The measured cross section is
\ifthenelse{\boolean{cms@external}}{
\begin{multline}
  \sigma (\pp \to \cPZ) \mathcal{B}(\cPZ \to \elfour) = \\
         249 \pm 11 \stat _{-15}^{+16} \syst \pm 4 \thy \pm 6 \lum \unit{fb}.
\label{xstot:z4l}
\end{multline}
}{
\begin{equation}
  \sigma (\pp \to \cPZ) \mathcal{B}(\cPZ \to \elfour) = 
         249 \pm 11 \stat _{-15}^{+16} \syst \pm 4 \thy \pm 6 \lum \unit{fb}.
\label{xstot:z4l}
\end{equation}
}
The branching fraction for the $\cPZ \to \elfour$ decay,
$\mathcal{B}(\cPZ \to \elfour)$,
is measured by comparing the cross section given by Eq.~(\ref{xstot:z4l})
with the $\cPZ \to \elpelm$ cross section, and is computed as
\ifthenelse{\boolean{cms@external}}{
\begin{multline}
\mathcal{B}(\cPZ \to \elfour) =\\
\frac{\sigma (\pp \to \cPZ \to \elfour)}
{\mathcal{C}^{\text{60--120}}_{\text{80--100}} \,
\sigma (\pp \to \cPZ \to \ell^+\ell^-) / \mathcal{B}(\cPZ \to \ell^+\ell^-)},
\end{multline}
}{
\begin{equation}
\mathcal{B}(\cPZ \to \elfour) =
\frac{\sigma (\pp \to \cPZ \to \elfour)}
{\mathcal{C}^{\text{60--120}}_{\text{80--100}} \,
\sigma (\pp \to \cPZ \to \ell^+\ell^-) / \mathcal{B}(\cPZ \to \ell^+\ell^-)},
\end{equation}}
where $\sigma (\pp \to \cPZ \to \elpelm) =
1870 _{-40}^{+50}\unit{pb} $ is the
$\cPZ  \to  \elpelm$ cross section times branching fraction
calculated at NNLO with \textsc{fewz}~v2.0~\cite{Gavin:2010az} in the mass range
60--120\GeV. Its uncertainty includes PDF
uncertainties and uncertainties in $\alpha_s$, the charm and bottom quark masses,
and the effect of neglected higher-order corrections to the calculation.
The factor $\mathcal{C}^{\text{60--120}}_{\text{80--100}} = 0.926 \pm 0.001$ corrects for the
difference in $\cPZ$ boson mass windows and
is estimated using \POWHEG. Its uncertainty includes scale and PDF
variations.
The nominal $\cPZ$ to dilepton branching fraction
$\mathcal{B}(\cPZ \to \elpelm)$
is 0.03366~\cite{Olive:2016xmw}. The measured value is
\ifthenelse{\boolean{cms@external}}{
\begin{multline}
  \mathcal{B}(\cPZ \to \elfour) =\\ 4.83 _{-0.22}^{+0.23} \stat _{-0.29}^{+0.32} \syst \pm 0.08 \thy \pm 0.12 \lum \times 10^{-6}, 
\end{multline}
}{
\begin{equation}
  \mathcal{B}(\cPZ \to \elfour) = 4.83 _{-0.22}^{+0.23} \stat _{-0.29}^{+0.32} \syst \pm 0.08 \thy \pm 0.12 \lum \times 10^{-6},
\end{equation}
}
where the theoretical uncertainty includes the uncertainties in
$\sigma (\pp \to \cPZ) \mathcal{B} (\cPZ \to \elpelm)$,
$\mathcal{C}^{\text{60--120}}_{\text{80--100}}$,
and $\mathcal{A}$.
This can be compared with $4.6 \times 10^{-6}$, computed with
\MGvATNLO,
and is consistent with the CMS and ATLAS measurements at
$\sqrt{s} = 7, 8,$ and 13\TeV~\cite{CMS:2012bw,Aad:2014wra,Khachatryan:2016txa}.

The total $\ZZ$ production cross section for both dileptons produced in the
mass range 60--120\GeV and $m_{\ell^+\ell^{\prime -}} > 4\GeV$ is found to be
\ifthenelse{\boolean{cms@external}}{
\begin{multline}
 \sigma(\pp \to \ZZ) =\\ 17.5 _{-0.5}^{+0.6} \stat \pm 0.6 \syst \pm 0.4 \thy \pm 0.4 \lum \unit{pb}.
\end{multline}
}{
\begin{equation}
  \sigma(\pp \to \ZZ) = 17.5 _{-0.5}^{+0.6} \stat \pm 0.6 \syst \pm 0.4 \thy \pm 0.4 \lum \unit{pb}.
\end{equation}
}
The measured total cross section can be compared to the theoretical value of
$14.5^{+0.5}_{-0.4} \pm 0.2\unit{pb}$ calculated
with a combination of \POWHEG and \MCFM with the same settings as described for
$\sigma_{\text{fid}} (\pp \to \ZZ \to \elfour)$.
It can also be compared to
$16.2^{+0.6}_{-0.4}$\unit{pb}, calculated at NNLO in QCD via
\textsc{matrix}~v1.0.0\_beta4~\cite{Cascioli:2014yka,Grazzini:2015hta}, or
$15.0^{+0.7}_{-0.6} \pm 0.2$\unit{pb}, calculated with \MCFM at NLO in QCD with
additional contributions from LO $\Pg\Pg  \to  \cPZ\cPZ$ diagrams. Both values are
calculated with the NNPDF3.0 PDF sets, at NNLO and NLO, respectively, and fixed
scales set to $\mu_\mathrm{F} = \mu_\mathrm{R} = m_\cPZ$.

{\tolerance=800
This measurement agrees with the previously published cross section measured by CMS
at 13\TeV~\cite{Khachatryan:2016txa} based on a 2.6\fbinv data sample collected in 2015:
\ifthenelse{\boolean{cms@external}}{
\begin{multline}
 \sigma(\pp \to \ZZ) =\\ 14.6 ^{+1.9}_{-1.8} \stat _{-0.5}^{+0.3} \syst \pm 0.2 \thy \pm 0.4 \lum \unit{pb}.
\end{multline}
}{
\begin{equation}
 \sigma(\pp \to \ZZ) = 14.6 ^{+1.9}_{-1.8} \stat _{-0.5}^{+0.3} \syst \pm 0.2 \thy \pm 0.4 \lum \unit{pb}.
\end{equation}
}
The two measurements can be combined to yield the ``2015+2016 cross section''
\ifthenelse{\boolean{cms@external}}{
\begin{multline}
 \sigma(\pp \to \ZZ) =\\ 17.2 \pm 0.5 \stat \pm 0.7 \syst \pm 0.4 \thy \pm 0.4 \lum \unit{pb}.
\end{multline}
}{
\begin{equation}
 \sigma(\pp \to \ZZ) = 17.2 \pm 0.5 \stat \pm 0.7 \syst \pm 0.4 \thy \pm 0.4 \lum \unit{pb}.
\end{equation}
}
The combination was performed once considering the experimental uncertainties
to be fully correlated between the 2015 and 2016 data sets, and once considering them
to be fully uncorrelated. The results were averaged, and the difference was added
linearly to the systematic uncertainty in the combined cross section.
\par}

\begin{figure}[htbp]
\centering
\includegraphics[width=\cmsFigWidth]{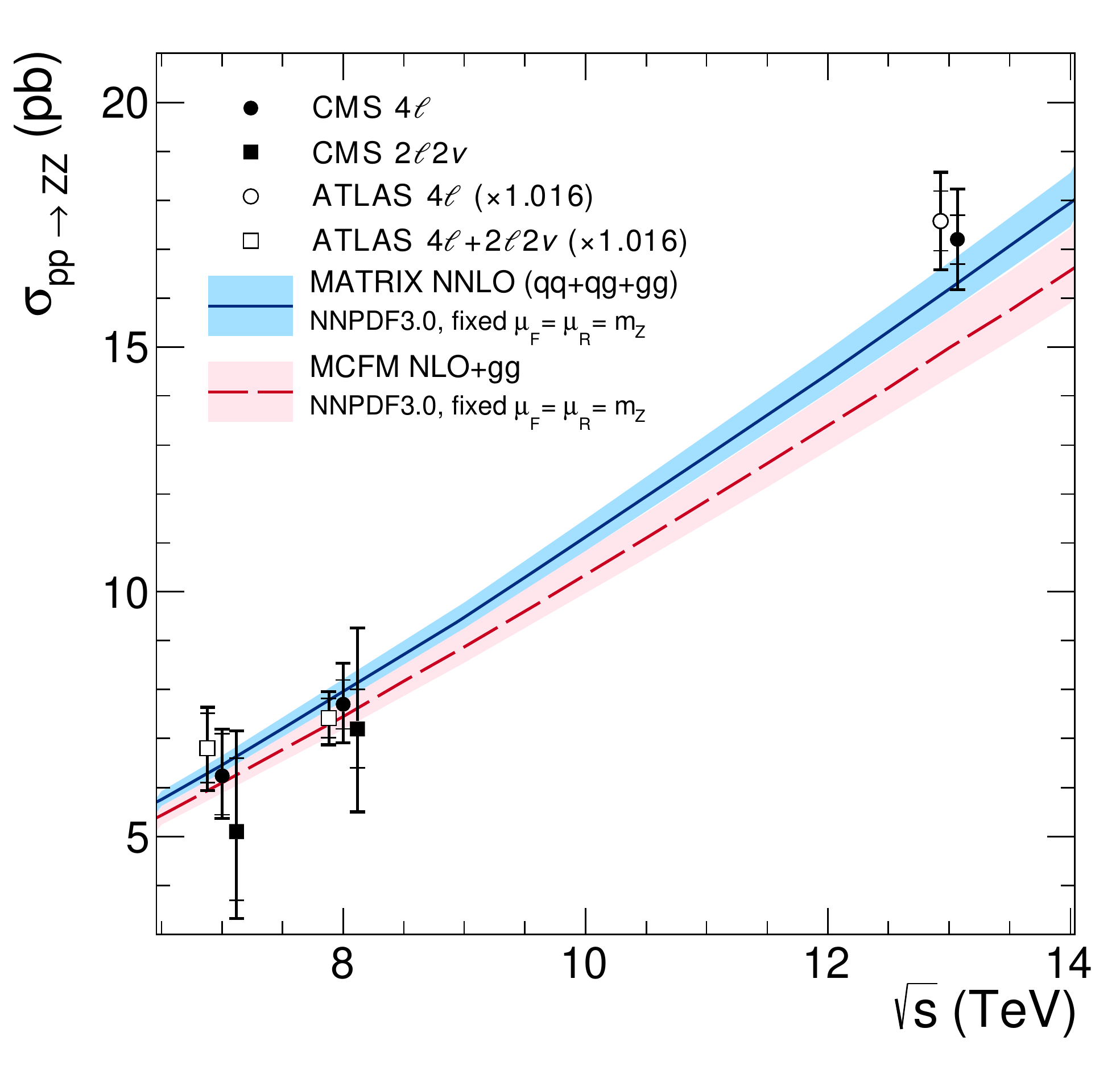}
\caption{
        The total ZZ cross section as a function of the proton-proton
        center-of-mass energy. Results from the CMS and ATLAS experiments
        are compared to predictions from \textsc{matrix} at NNLO in QCD, and \MCFM at NLO
        in QCD. The \MCFM prediction also includes gluon-gluon initiated
        production at LO in QCD. Both predictions use NNPDF3.0 PDF sets and fixed scales
        $\mu_\mathrm{F} = \mu_\mathrm{R} = m_\cPZ$. Details of the calculations and uncertainties are given
        in the text. The ATLAS measurements were performed with a $\cPZ$ boson
        mass window of 66--116\GeV, and are corrected for the resulting 1.6\% difference.
        Measurements at the same center-of-mass energy are shifted slightly along the
        horizontal axis for clarity.
}
\label{fig:xsec_vs_sqrts}
\end{figure}

The total $\ZZ$ cross section is shown
in Fig.~\ref{fig:xsec_vs_sqrts} as a function of the proton-proton
center-of-mass energy. Results from
CMS~\cite{Chatrchyan:2012sga, CMS:2014xja}
and ATLAS~\cite{Aad:2012awa,Aad:2015rka,Aaboud:2017rwm}
are compared to predictions from \textsc{matrix} and \MCFM with
the NNPDF3.0 PDF sets and fixed scales $\mu_\mathrm{F} = \mu_\mathrm{R} = m_\cPZ$.
The \textsc{matrix} prediction uses PDFs calculated at NNLO, while the
\MCFM prediction uses NLO PDFs.
The uncertainties are statistical (inner bars) and
statistical and systematic added in quadrature (outer bars). The band
around the \textsc{matrix} predictions reflects scale uncertainties, while
the band around the \MCFM predictions reflects both scale and PDF
uncertainties.

The measurement of the differential cross sections
provides detailed information about $\cPZ\cPZ$ kinematics.
The observed yields are unfolded using the iterative technique described
in Ref.~\cite{unfolding}. Unfolding is performed with the RooUnfold
package~\cite{Adye:2011gm} and regularized by stopping after four iterations.
Statistical uncertainties in the data distributions are propagated through the
unfolding process to give the statistical uncertainties on the normalized
differential cross sections.

The three decay channels, $4\Pe$, $4\Pgm$,  and $2\Pe 2\Pgm$, are
combined after unfolding because no differences are expected in their kinematic distributions.
The generator-level leptons used for the unfolding are dressed as in the
fiducial cross section calculation.

\begin{figure*}[htbp]
\centering
\includegraphics[width=0.49\textwidth]{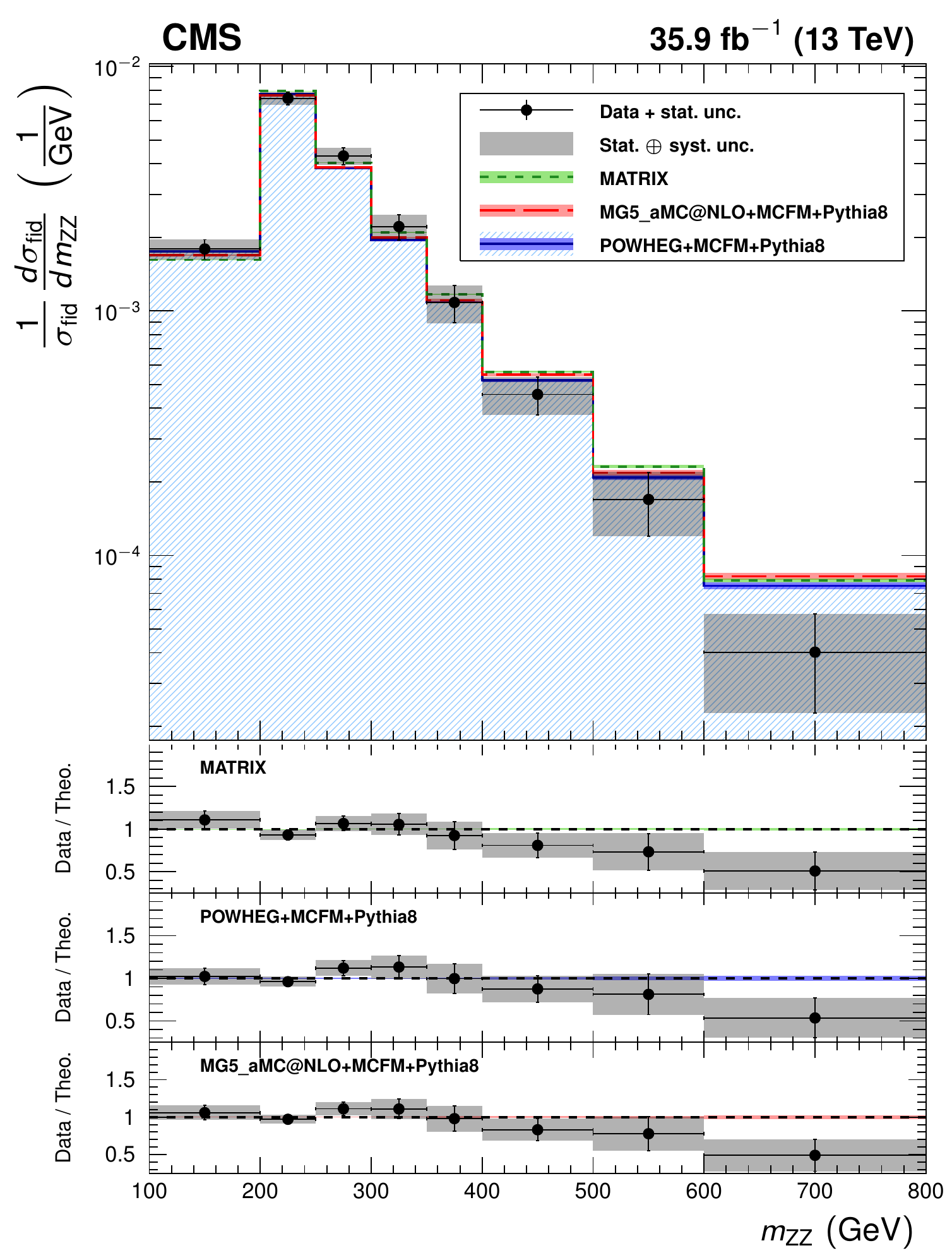}
\includegraphics[width=0.49\textwidth]{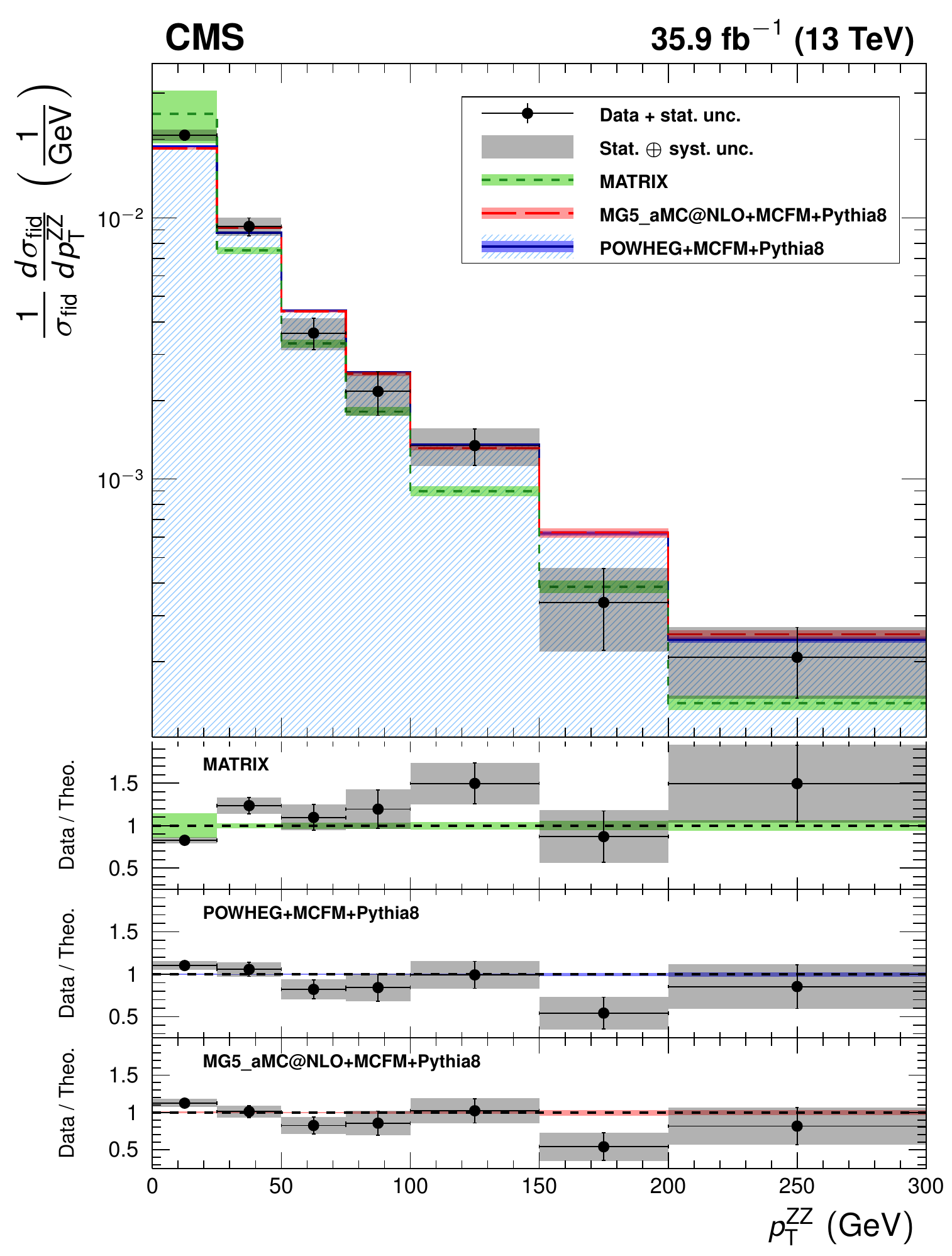}
\caption{
Differential cross sections normalized to the fiducial cross section
for the combined 4$\Pe$, 4$\Pgm$, and 2$\Pe$2$\Pgm$ decay channels as a function of
mass (left) and $\pt$ (right) of the {\ZZ} system.
Points represent the unfolded data; the solid, dashed, and dotted histograms represent
the {\POWHEG}+{\MCFM}, \MGvATNLO+{\MCFM}, and \textsc{matrix}
predictions for $\PZ\PZ$ signal, respectively, and the bands around
the predictions reflect their combined statistical, scale, and PDF uncertainties.
\PYTHIA~v8 was used for parton showering, hadronization, and underlying event
simulation in the \POWHEG, \MGvATNLO, and {\MCFM} samples.
The lower part of each plot represents the ratio of the measured cross section
to the theoretical distributions. The shaded grey
areas around the points represent the sum in quadrature
of the statistical and systematic uncertainties, while the crosses represent the statistical
uncertainties only.
}
\label{fig:diff1}
\end{figure*}

\begin{figure}[htbp]
\centering
\includegraphics[width=0.49\textwidth]{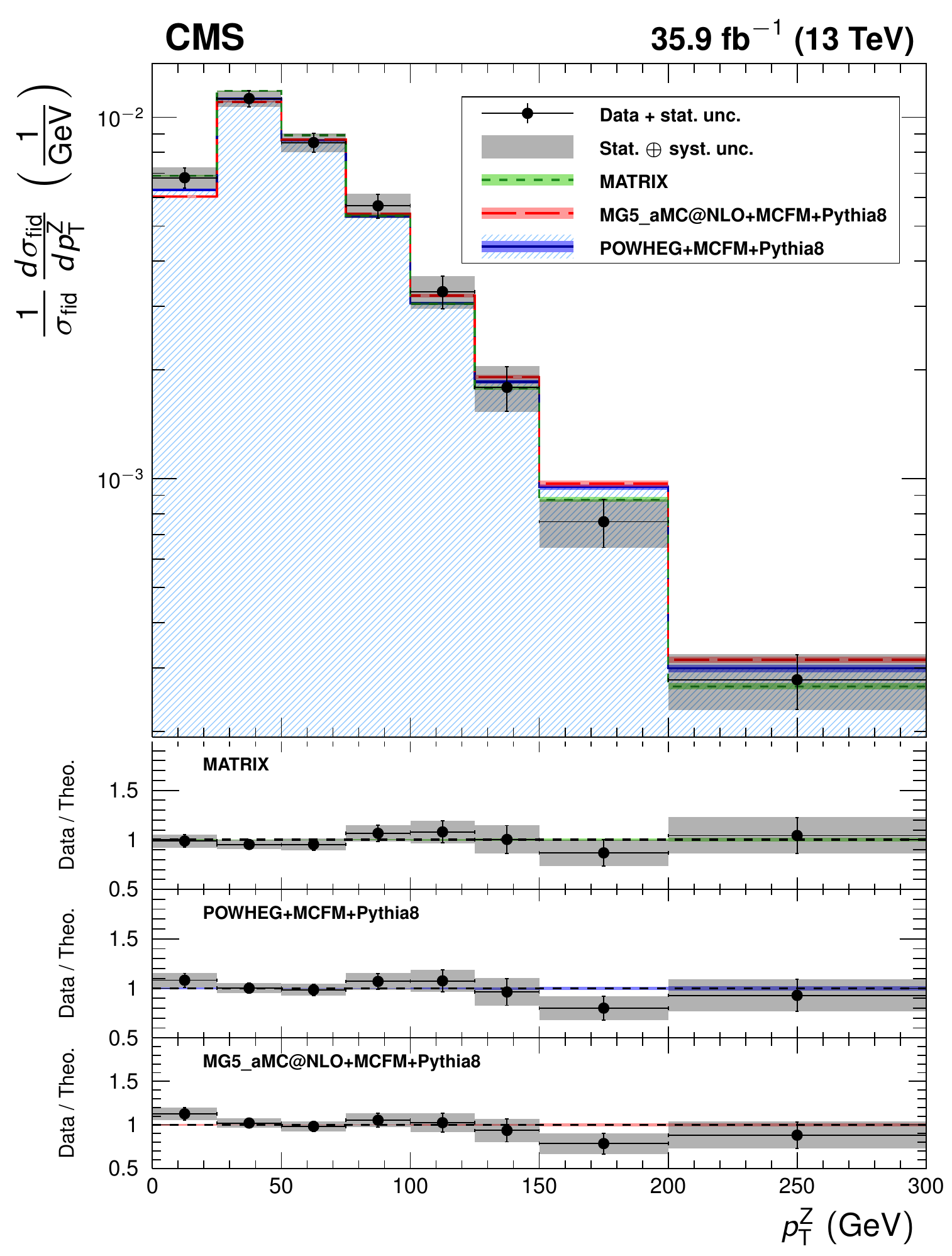}
\includegraphics[width=0.49\textwidth]{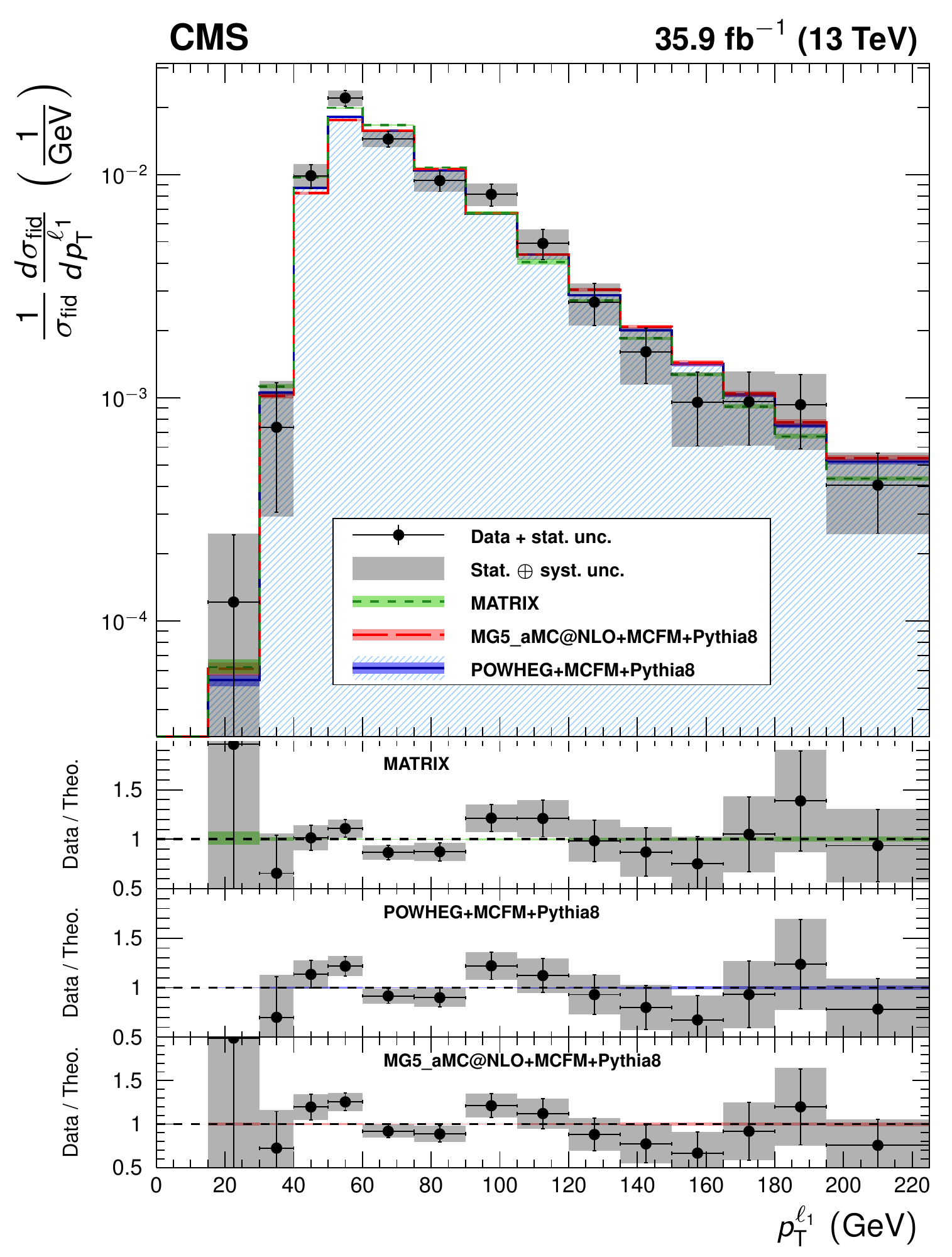}
\caption{
Normalized {\ZZ} differential cross sections as a function of
the $\pt$ of (\cmsLeft) all $\cPZ$ bosons
and (\cmsRight) the leading lepton in $\ZZ$ events.
Other details are as described in the caption of Fig.~\ref{fig:diff1}.
}
\label{fig:diff1a}
\end{figure}

\begin{figure}[htbp]
\centering
\includegraphics[width=0.49\textwidth]{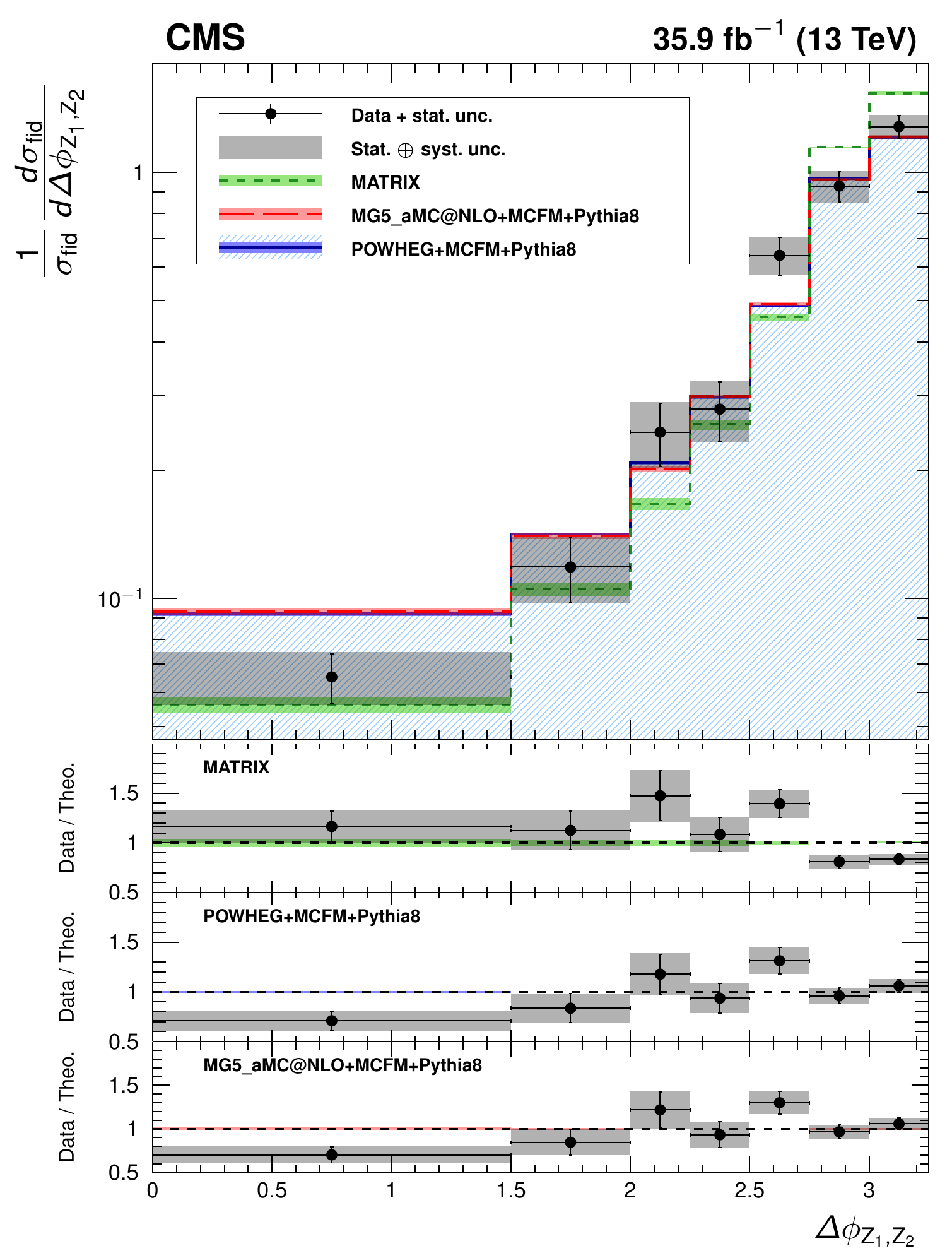}
\includegraphics[width=0.49\textwidth]{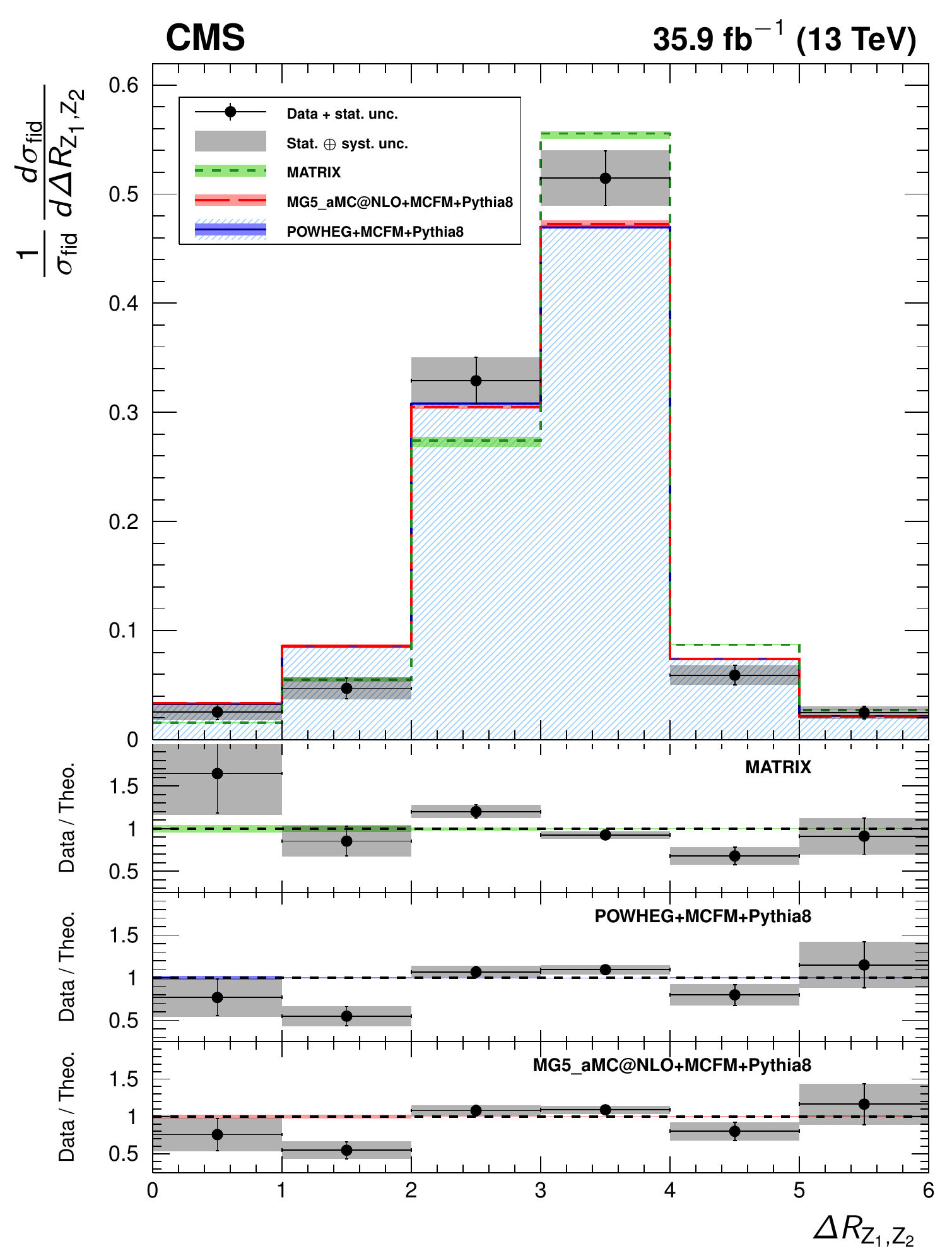}
\caption{
Normalized {\ZZ} differential cross sections as a function of
(\cmsLeft) the azimuthal separation of the two $\cPZ$ bosons and
(\cmsRight) $\Delta R$ between the $\cPZ$-bosons.
Other details are as described in the caption of Fig.~\ref{fig:diff1}.
}
\label{fig:diff2}
\end{figure}

The differential distributions normalized to the fiducial cross sections are
presented
in Figs.~\ref{fig:diff1}--\ref{fig:diff2} for the combination
of the 4$\Pe$, 4$\Pgm$, and 2$\Pe$2$\Pgm$
decay channels. The fiducial cross section definition includes $\pt^{\ell}$ and
$\abs{\eta^{\ell}}$ selections on each lepton,
and the 60--120\GeV mass requirement, as described in Table~\ref{table:fiducial_cuts}
and Section~\ref{sec:eventreconstruction}.
Figure~\ref{fig:diff1} shows the normalized differential cross sections as
functions of the mass and $\pt$ of the {\ZZ} system,
Fig.~\ref{fig:diff1a} shows them as functions of the {\pt} of all {\cPZ} bosons
and the {\pt} of the leading lepton in each event, and
Fig.~\ref{fig:diff2} shows the angular correlations between
the two $\cPZ$ bosons.
The data are corrected for background contributions and compared with
the theoretical predictions from \POWHEG and \MCFM, \MGvATNLO
and \MCFM, and \textsc{matrix}. The bottom part of each plot shows
the ratio of the measured to the predicted values. The bin sizes are chosen
according to the resolution of the relevant variables, while also keeping
the statistical uncertainties at a similar level in all bins. The data are
well reproduced by the simulation except in the low $\pt$ regions, where data
tend to have a steeper slope than the prediction.

\begin{figure}[htbp]
  \centering
  \includegraphics[width=\cmsFigWidth]{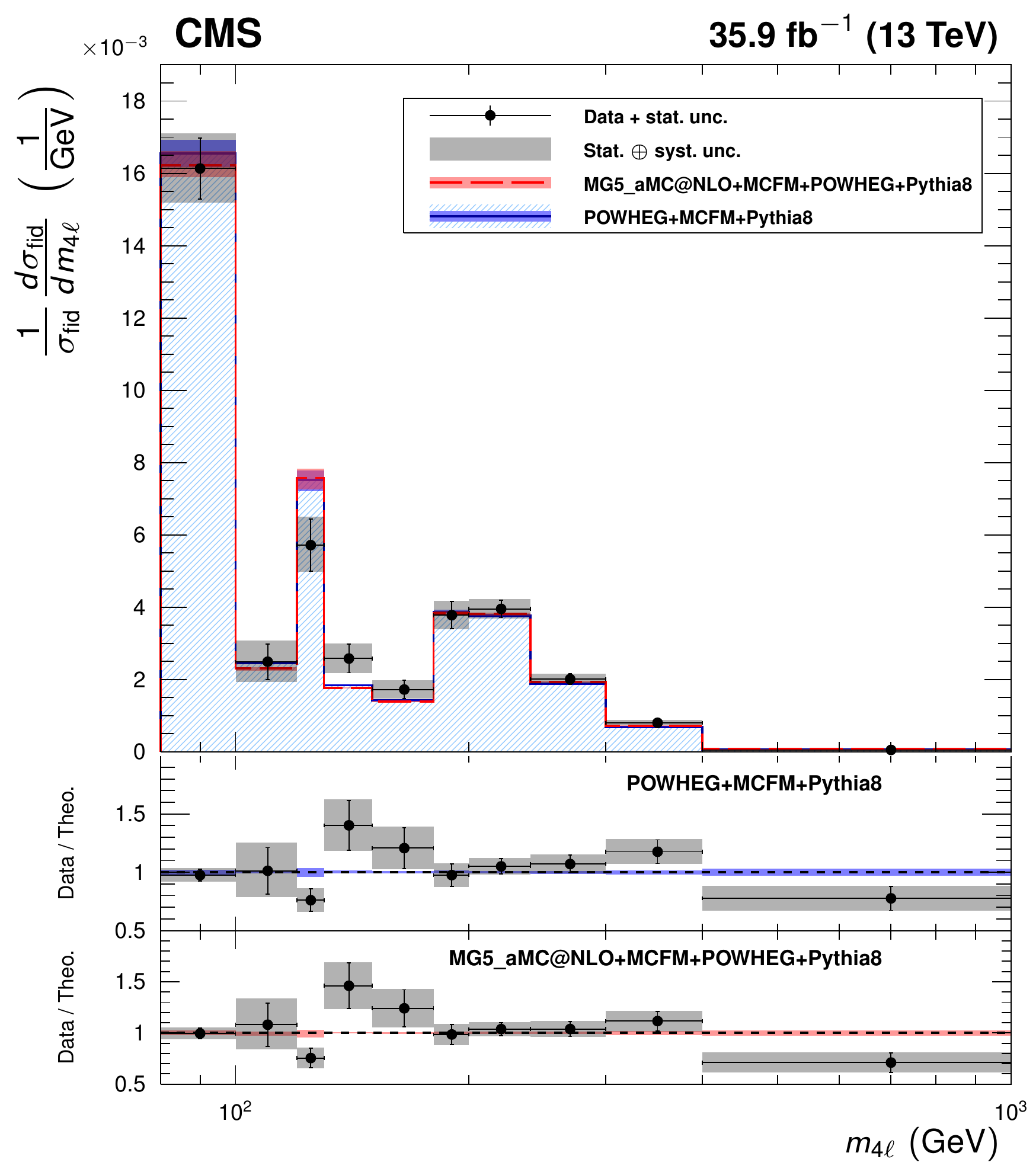}
  \caption{
The normalized differential four-lepton cross section as a function of the
four-lepton mass, subject only to the common requirements of
Table~\ref{table:fiducial_cuts}. SM $\Pg\Pg \to \PH \to \cPZ\cPZ^\ast$
production is included, simulated with \POWHEG.
Other details are as described in the caption of Fig.~\ref{fig:diff1}.
}
\label{fig:unfold_full}
\end{figure}

Figure~\ref{fig:unfold_full} shows the normalized differential four-lepton cross
section as a function of $m_{4\ell}$, subject only to the
common requirements of Table~\ref{table:fiducial_cuts}. This includes
contributions from the \cPZ\ and Higgs boson resonances and continuum
\cPZ\cPZ\ production.

\section{Limits on anomalous triple gauge couplings}

\begin{figure}[htbp]
\centering
\includegraphics[width=\cmsFigWidth]{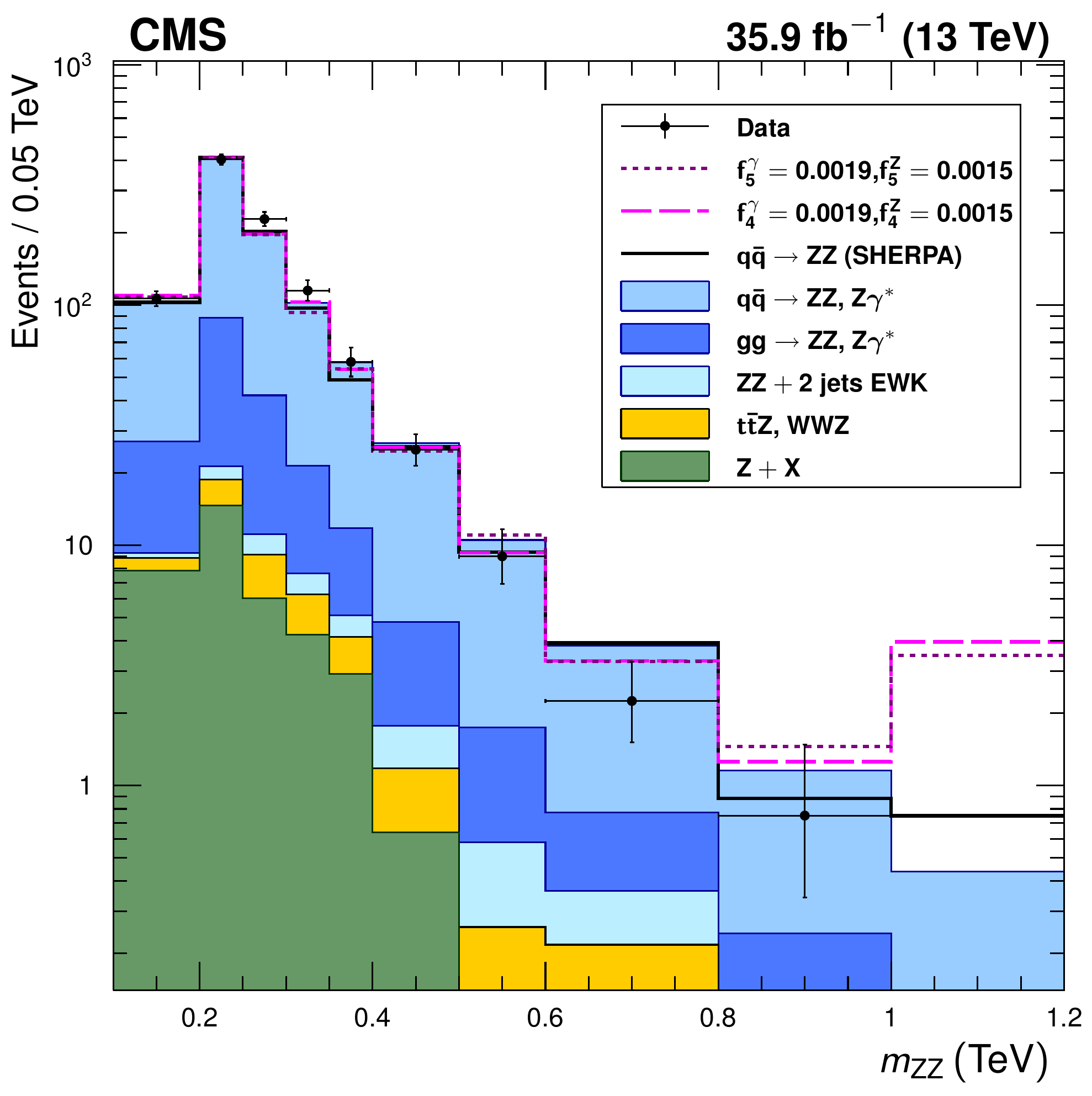}
\caption{
 Distribution of the four-lepton reconstructed mass for the combined
$4\Pe$, 4$\Pgm$, and $2\Pe2\Pgm$ channels.
Points represent the data, the filled histograms represent the SM
expected yield including signal and irreducible background predictions from
simulation and the
data-driven background estimate. Unfilled histograms represent examples of aTGC
signal predictions (dashed), and the \SHERPA SM prediction (solid), included to illustrate
the expected shape differences between the \SHERPA and \POWHEG predictions.
Vertical bars on the data points show their statistical uncertainty.
The \SHERPA distributions are normalized such that the SM sample has the same
total yield as the \POWHEG sample predicts. Bin contents
are normalized to the bin widths, using a unit bin size of 50\GeV; horizontal
bars on the data points show the range of the corresponding bin.
The last bin includes the ``overflow''
contribution from events at masses above 1.2\TeV.
}
\label{figure:sherpa4l}
\end{figure}

The presence of aTGCs
would increase the yield of events at high four-lepton masses.
Figure~\ref{figure:sherpa4l} presents the
distribution of the four-lepton reconstructed mass of events with both {\cPZ}
bosons in the mass range 60--120\GeV for the combined
$4\Pe$, 4$\Pgm$, and $2\Pe2\Pgm$ channels. This distribution
is used to set the limits on possible contributions from aTGCs. Two simulated
samples with nonzero aTGCs are shown as examples, along with the SM
distribution simulated by both \SHERPA and \POWHEG.

\begin{figure}[htbp]
\centering
\includegraphics[width=0.48\textwidth]{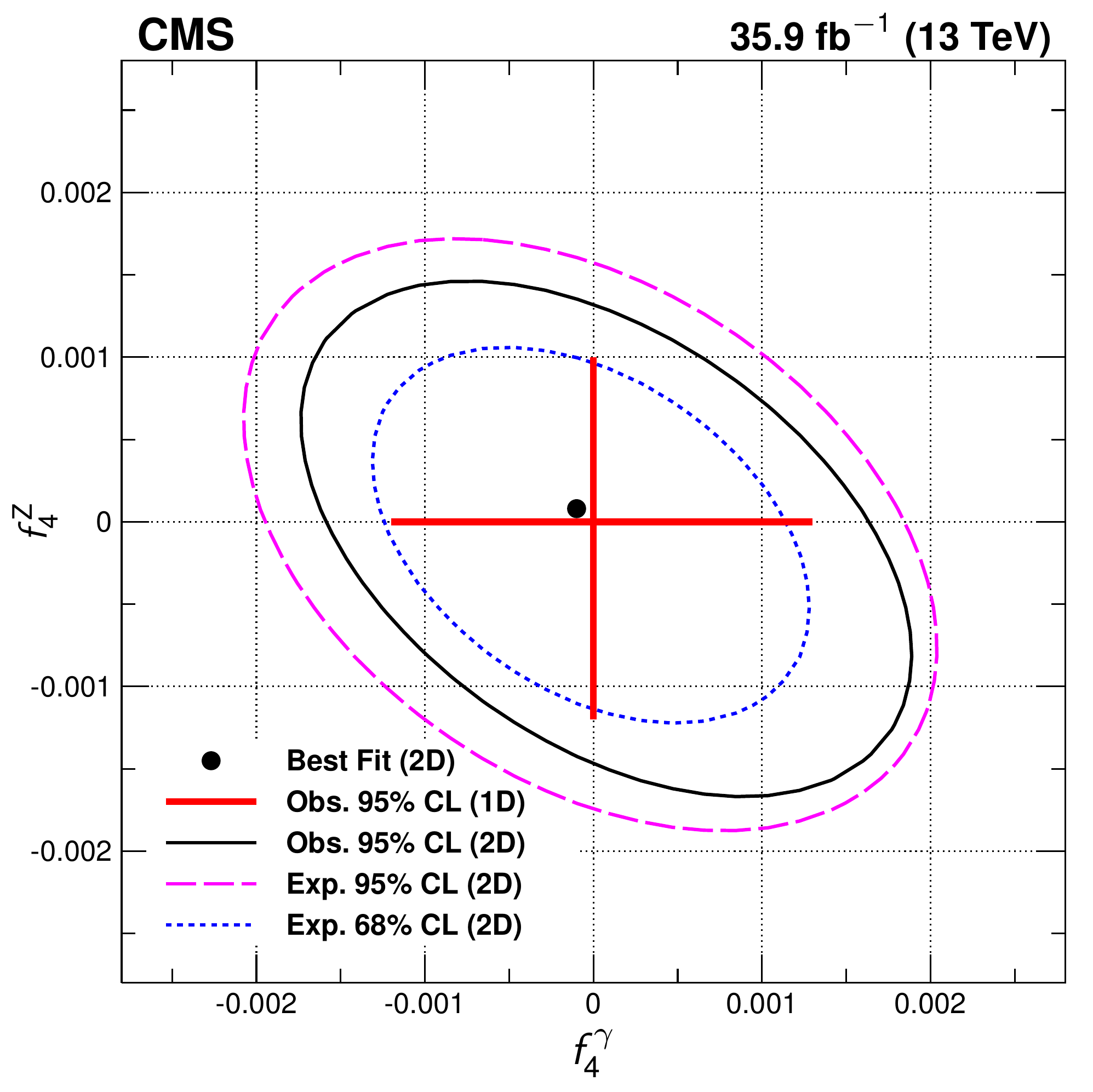}
\includegraphics[width=0.48\textwidth]{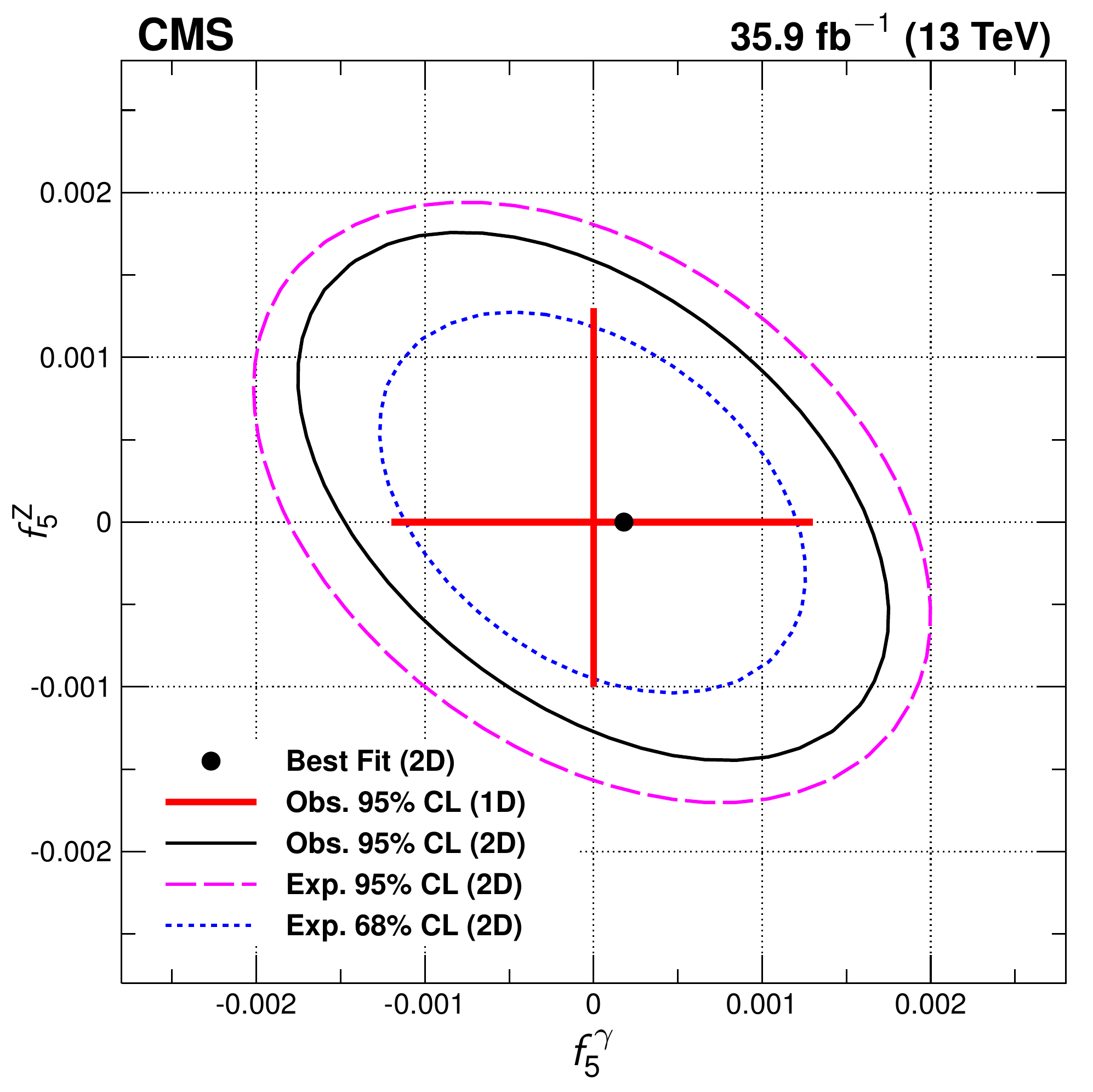}
\caption{Two-dimensional observed 95\% CL limits (solid contour) and expected
68 and 95\% CL limits (dashed contour) on the $\cPZ\cPZ\cPZ$
and $\cPZ\cPZ\gamma$ aTGCs. The
\cmsLeft (\cmsRight) plot shows the exclusion contour in the
$f_{4(5)}^\cPZ, f_{4(5)}^\gamma$ parameter planes. The values of couplings
outside of contours are excluded at the corresponding confidence level.
The solid dot is the point at which
the likelihood is at its maximum.
The solid lines at the center show the observed one-dimensional 95\% CL limits
for $f_{4,5}^\gamma$ (horizontal) and $f_{4,5}^\cPZ$ (vertical).
No form factor is used. }
\label{figure:aTGC}
\end{figure}

The invariant mass distributions are interpolated from the
\SHERPA simulations for different values of the anomalous couplings in the range between 0 and 0.015. For
each distribution, only one or two couplings are varied while all others are set to zero.
The measured signal is obtained from a comparison of the data to
a grid of aTGC models in the
$(f_{4}^\cPZ, f_{4}^\gamma)$ and
$(f_{5}^\cPZ, f_{5}^\gamma)$
parameter
planes.  Expected signal values are interpolated between the 2D grid
points using a second-degree polynomial, since the cross
section for the signal depends quadratically on the coupling parameters.
A binned profile likelihood method, Wald Gaussian approximation, and Wilk's theorem are used
to derive one-dimensional limits at a 95\% confidence level (CL) on each
of the four aTGC parameters, and two-dimensional limits at a 95\% CL on the pairs ($f_4^\cPZ$, $f_4^\gamma$) and
($f_5^\cPZ$, $f_5^\gamma$)~\cite{Olive:2016xmw,Wilks:1938dza,Cowan:2010js}.
When the limits are calculated for each parameter or pair, all other parameters are set
to their SM values.
The systematic uncertainties described in Section~\ref{sec:systematics}
are treated as nuisance parameters with log-normal distributions.
No form factor is used when deriving the limits so that the
results do not depend  on any assumed energy scale characterizing new physics.
The constraints on anomalous couplings are displayed in Fig.~\ref{figure:aTGC}.
The curves indicate 68 and 95\% confidence levels, and
the solid dot shows the coordinates where the likelihood reaches
its maximum. Coupling values outside the contours
are excluded at the corresponding confidence levels.
The limits are dominated by statistical uncertainties.

The observed one-dimensional 95\% CL limits
for the $f_4^{\cPZ,\gamma}$ and $f_5^{\cPZ,\gamma}$ anomalous coupling parameters are:
\ifthenelse{\boolean{cms@external}}{
\begin{equation}
  \begin{aligned}
  -0.0012 &<f_4^\cPZ    < 0.0010 ,\\
  -0.0010 &<f_5^\cPZ    < 0.0013 ,\\
  -0.0012 &<f_4^{\gamma}< 0.0013 ,\\
  -0.0012 &<f_5^{\gamma}< 0.0013 .
  \end{aligned}
\end{equation}
}{
\begin{equation}
  \begin{aligned}
  -0.0012 <f_4^\cPZ    < 0.0010 ,\quad -0.0010 <f_5^\cPZ    < 0.0013 , \\
  \quad -0.0012 <f_4^{\gamma}< 0.0013 ,\quad -0.0012 <f_5^{\gamma}< 0.0013 .
  \end{aligned}
\end{equation}
}
These are the most stringent limits to date on
anomalous $\cPZ\cPZ\cPZ$ and $\cPZ\cPZ\gamma$ trilinear gauge boson
couplings, improving on the previous strictest
results from CMS~\cite{Khachatryan:2015pba} by factors of two or more
and constraining the coupling parameters more than the
corresponding ATLAS results~\cite{Aaboud:2017rwm}.

\begin{figure}[htbp]
\centering
\includegraphics[width=0.45\textwidth]{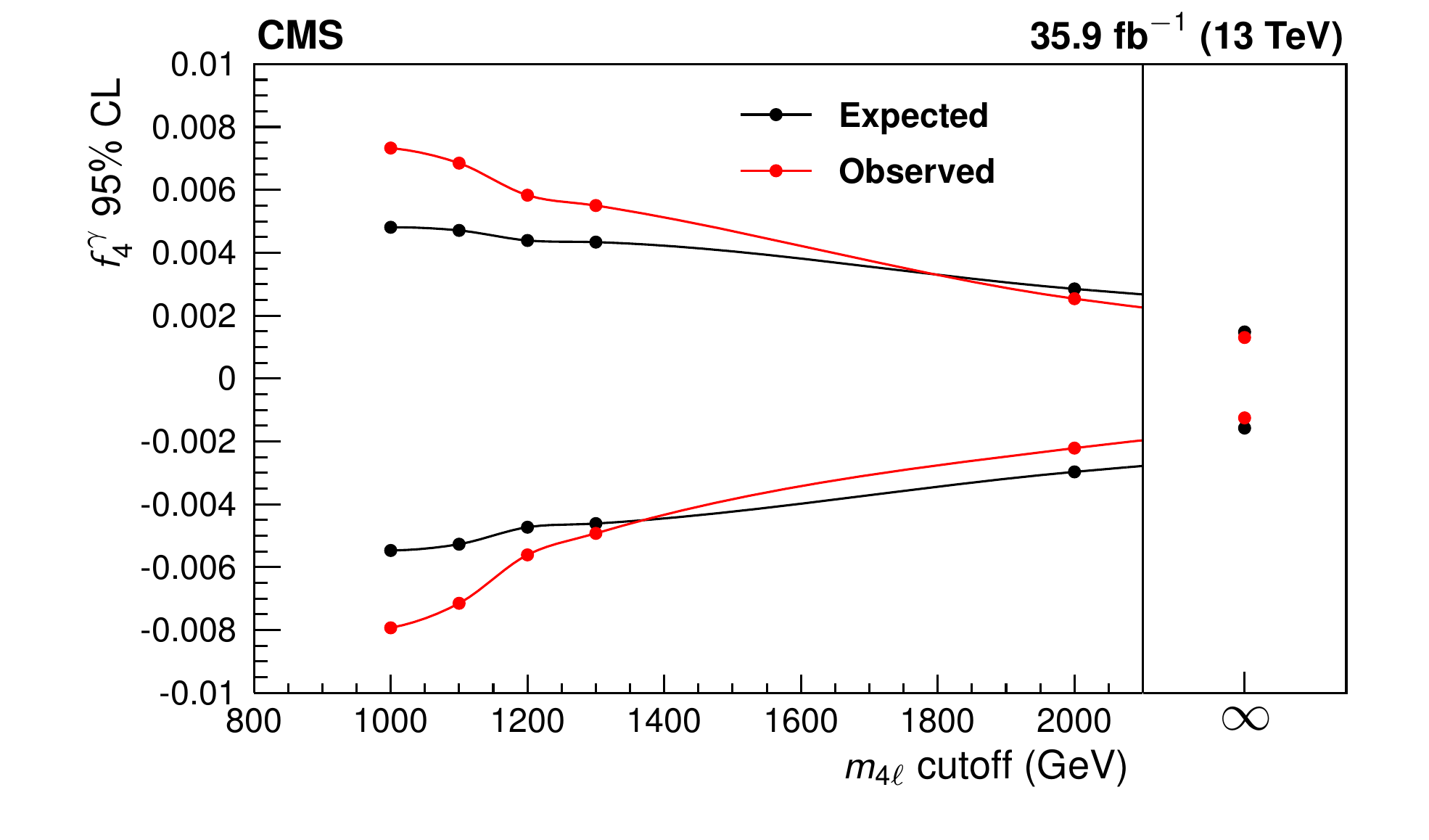}
\includegraphics[width=0.45\textwidth]{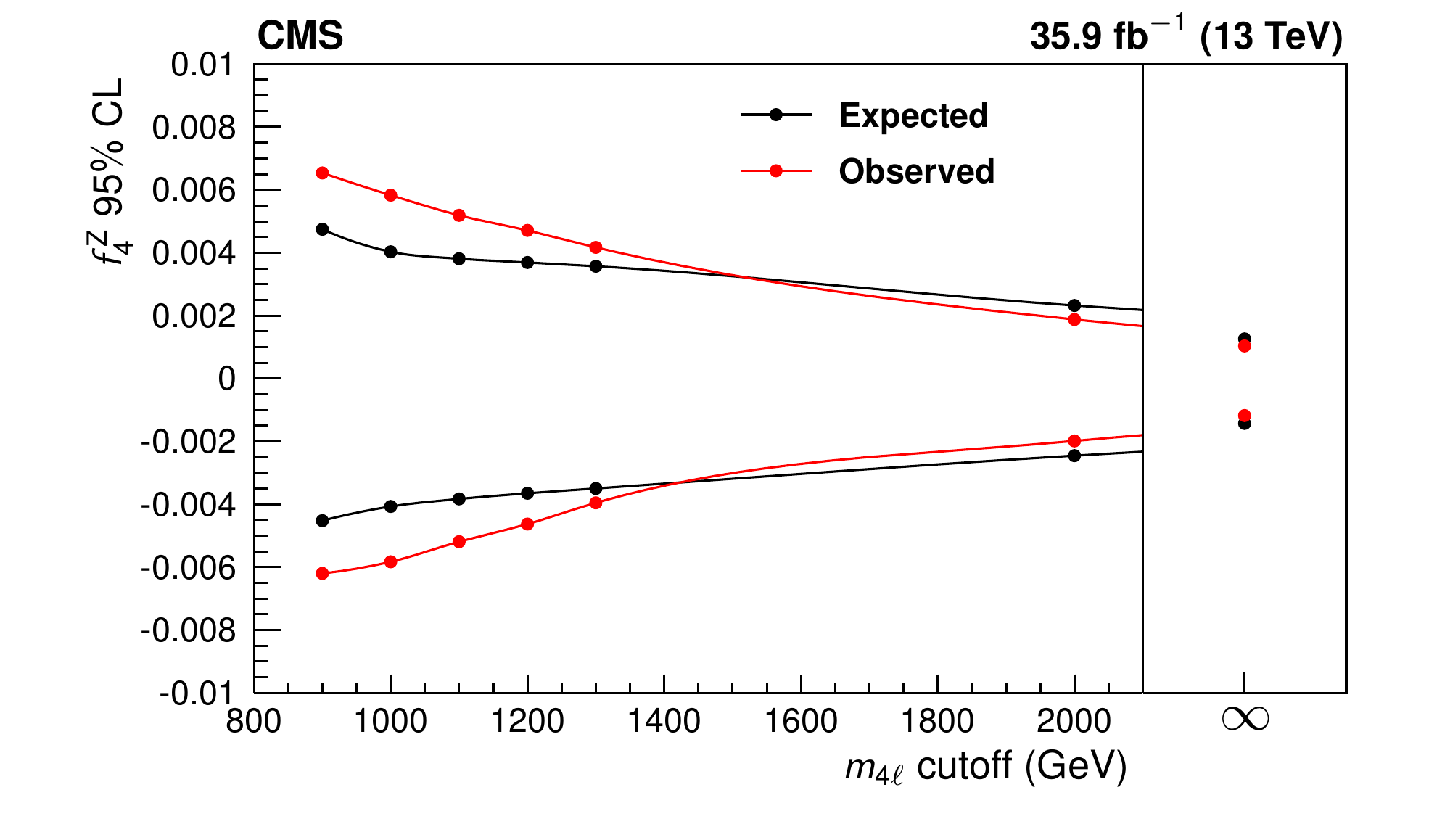}
\includegraphics[width=0.45\textwidth]{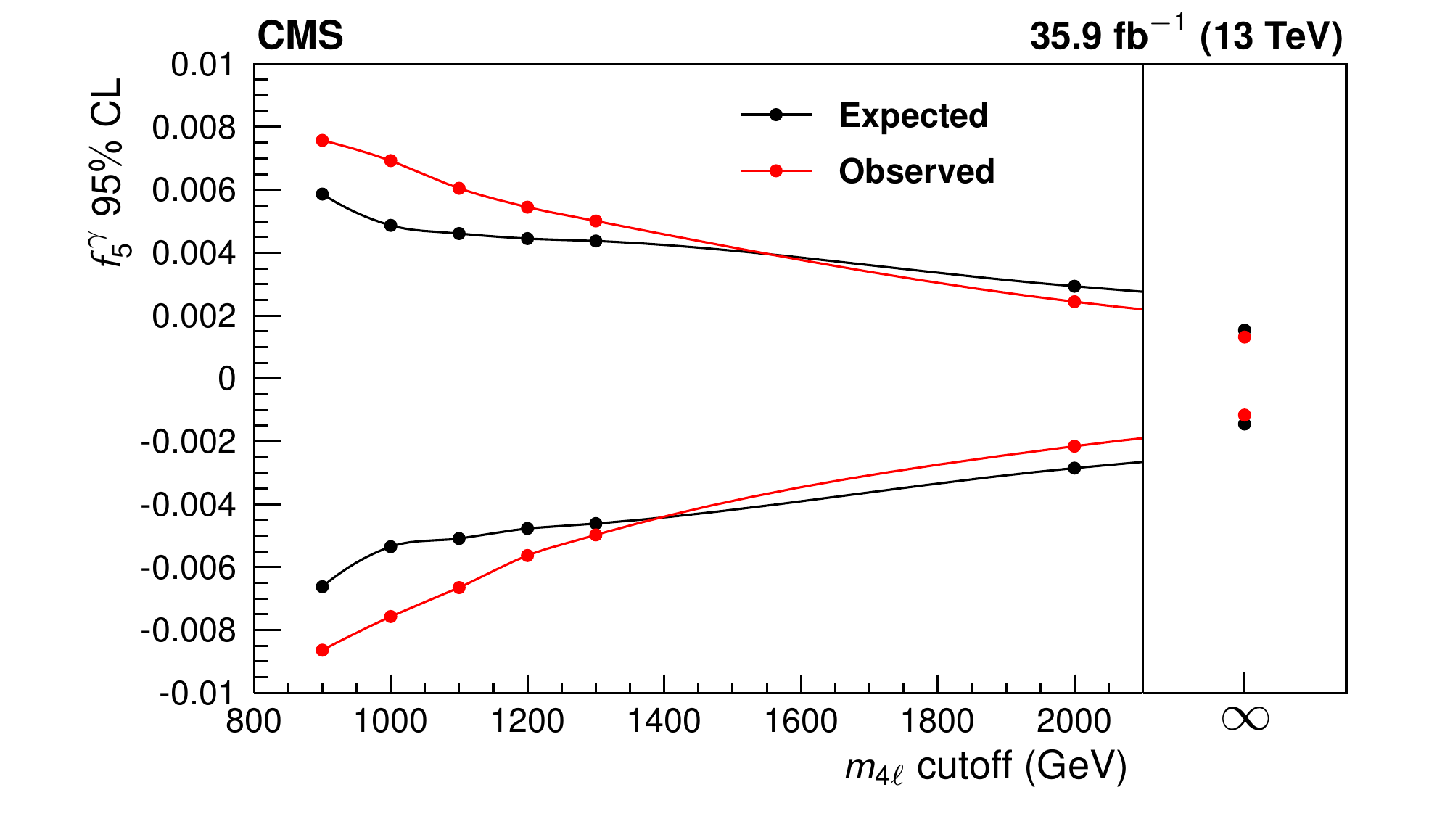}
\includegraphics[width=0.45\textwidth]{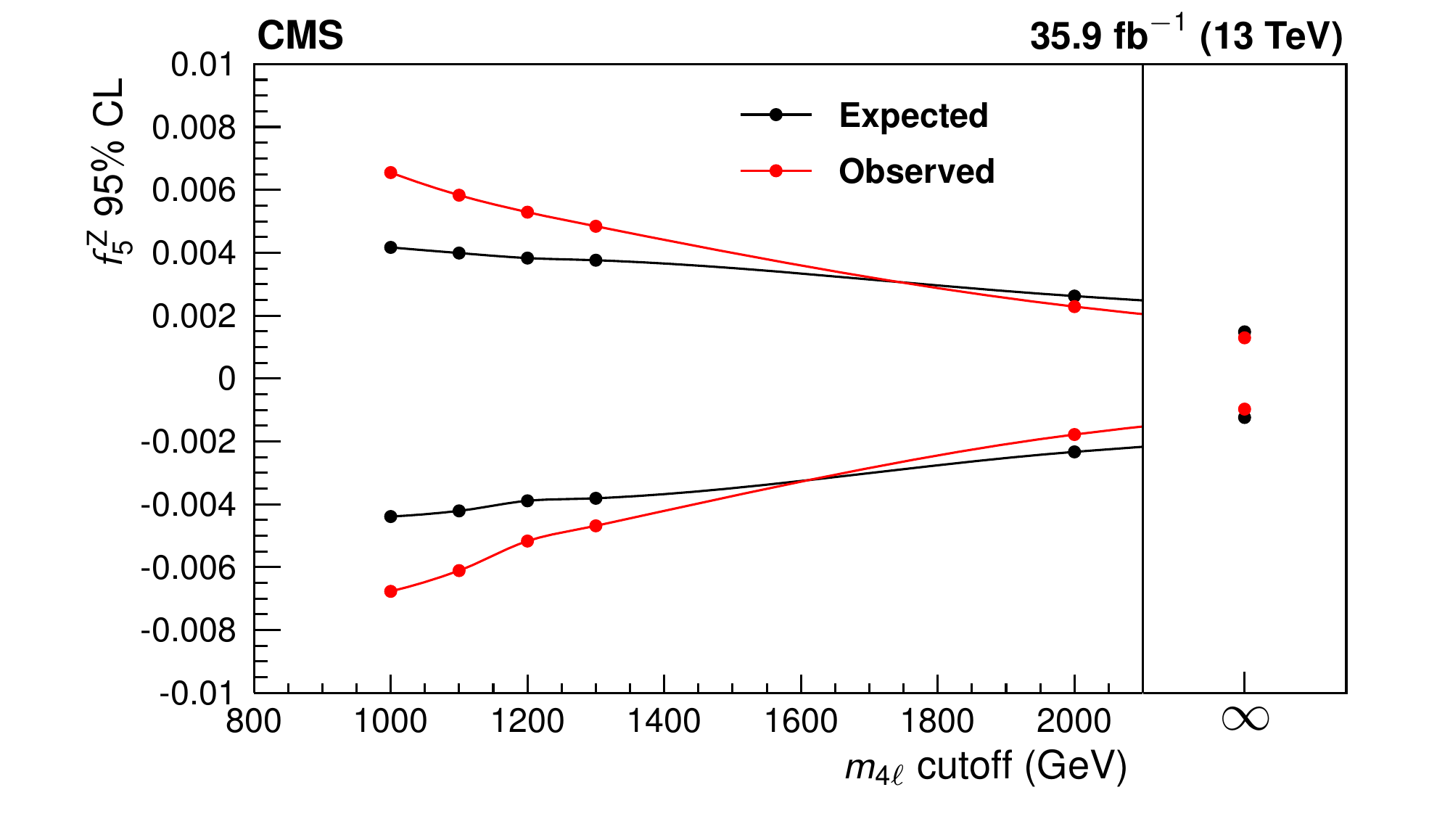}
\caption{
Expected and observed one-dimensional limits on the four aTGC
parameters, as a function of an upper cutoff on the invariant mass of the
four-lepton system. No form factor is used.
}
\label{figure:aTGCacutoff}
\end{figure}

One way to impose unitarity on the aTGC models is to restrict the range of
four-lepton invariant mass used in the limit calculation. The limits will then
depend on the ``cutoff'' value used. The computation of the one-dimensional
limits is repeated for different maximum allowed values of $m_{4\ell}$, and the
results are presented in Fig.~\ref{figure:aTGCacutoff} as a function of this
cutoff.

\section{Summary}

A series of measurements of four-lepton final states in
proton-proton collisions at $\sqrt{s} = 13\TeV$ have
been performed with the CMS detector at the
CERN LHC. The measured $\pp \to \ZZ$ cross section is
$\sigma(\Pp\Pp \to \cPZ\cPZ) = 17.2 \pm 0.5 \stat \pm 0.7 \syst \pm 0.4 \thy \pm 0.4 \lum \unit{pb}$
for $\cPZ$ boson masses in the range
$60 < m_{\cPZ} < 120\GeV$. The measured branching fraction for
$\cPZ$ boson decays to four leptons is
$\mathcal{B}(\cPZ \to \elfour) = 4.83 _{-0.22}^{+0.23} \stat _{-0.29}^{+0.32} \syst \pm 0.08 \thy \pm 0.12 \lum \times 10^{-6}$
for four-lepton mass in the range
$80 < m_{\elfour} < 100\GeV$ and dilepton mass
$m_{\ell\ell} > 4\GeV$ for all oppositely charged same-flavor lepton pairs.
Normalized differential cross sections were also measured.
All results agree
well with the SM predictions. Improved limits on
anomalous $\PZ \PZ \PZ$ and $\PZ \PZ \gamma$ triple gauge couplings
were established, the most stringent to date.

\begin{acknowledgments}
We congratulate our colleagues in the CERN accelerator departments for the excellent performance of the LHC and thank the technical and administrative staffs at CERN and at other CMS institutes for their contributions to the success of the CMS effort. In addition, we gratefully acknowledge the computing centers and personnel of the Worldwide LHC Computing Grid for delivering so effectively the computing infrastructure essential to our analyses. Finally, we acknowledge the enduring support for the construction and operation of the LHC and the CMS detector provided by the following funding agencies: BMWFW and FWF (Austria); FNRS and FWO (Belgium); CNPq, CAPES, FAPERJ, and FAPESP (Brazil); MES (Bulgaria); CERN; CAS, MoST, and NSFC (China); COLCIENCIAS (Colombia); MSES and CSF (Croatia); RPF (Cyprus); SENESCYT (Ecuador); MoER, ERC IUT, and ERDF (Estonia); Academy of Finland, MEC, and HIP (Finland); CEA and CNRS/IN2P3 (France); BMBF, DFG, and HGF (Germany); GSRT (Greece); OTKA and NIH (Hungary); DAE and DST (India); IPM (Iran); SFI (Ireland); INFN (Italy); MSIP and NRF (Republic of Korea); LAS (Lithuania); MOE and UM (Malaysia); BUAP, CINVESTAV, CONACYT, LNS, SEP, and UASLP-FAI (Mexico); MBIE (New Zealand); PAEC (Pakistan); MSHE and NSC (Poland); FCT (Portugal); JINR (Dubna); MON, RosAtom, RAS, RFBR and RAEP (Russia); MESTD (Serbia); SEIDI, CPAN, PCTI and FEDER (Spain); Swiss Funding Agencies (Switzerland); MST (Taipei); ThEPCenter, IPST, STAR, and NSTDA (Thailand); TUBITAK and TAEK (Turkey); NASU and SFFR (Ukraine); STFC (United Kingdom); DOE and NSF (USA).

\hyphenation{Rachada-pisek} Individuals have received support from the Marie-Curie program and the European Research Council and Horizon 2020 Grant, contract No. 675440 (European Union); the Leventis Foundation; the A. P. Sloan Foundation; the Alexander von Humboldt Foundation; the Belgian Federal Science Policy Office; the Fonds pour la Formation \`a la Recherche dans l'Industrie et dans l'Agriculture (FRIA-Belgium); the Agentschap voor Innovatie door Wetenschap en Technologie (IWT-Belgium); the Ministry of Education, Youth and Sports (MEYS) of the Czech Republic; the Council of Science and Industrial Research, India; the HOMING PLUS program of the Foundation for Polish Science, cofinanced from European Union, Regional Development Fund, the Mobility Plus program of the Ministry of Science and Higher Education, the National Science Center (Poland), contracts Harmonia 2014/14/M/ST2/00428, Opus 2014/13/B/ST2/02543, 2014/15/B/ST2/03998, and 2015/19/B/ST2/02861, Sonata-bis 2012/07/E/ST2/01406; the National Priorities Research Program by Qatar National Research Fund; the Programa Severo Ochoa del Principado de Asturias; the Thalis and Aristeia programs cofinanced by EU-ESF and the Greek NSRF; the Rachadapisek Sompot Fund for Postdoctoral Fellowship, Chulalongkorn University and the Chulalongkorn Academic into Its 2nd Century Project Advancement Project (Thailand); the Welch Foundation, contract C-1845; and the Weston Havens Foundation (USA). \end{acknowledgments}

\bibliography{auto_generated}

\cleardoublepage \appendix\section{The CMS Collaboration \label{app:collab}}\begin{sloppypar}\hyphenpenalty=5000\widowpenalty=500\clubpenalty=5000\textbf{Yerevan Physics Institute,  Yerevan,  Armenia}\\*[0pt]
A.M.~Sirunyan, A.~Tumasyan
\vskip\cmsinstskip
\textbf{Institut f\"{u}r Hochenergiephysik,  Wien,  Austria}\\*[0pt]
W.~Adam, F.~Ambrogi, E.~Asilar, T.~Bergauer, J.~Brandstetter, E.~Brondolin, M.~Dragicevic, J.~Er\"{o}, M.~Flechl, M.~Friedl, R.~Fr\"{u}hwirth\cmsAuthorMark{1}, V.M.~Ghete, J.~Grossmann, J.~Hrubec, M.~Jeitler\cmsAuthorMark{1}, A.~K\"{o}nig, N.~Krammer, I.~Kr\"{a}tschmer, D.~Liko, T.~Madlener, I.~Mikulec, E.~Pree, D.~Rabady, N.~Rad, H.~Rohringer, J.~Schieck\cmsAuthorMark{1}, R.~Sch\"{o}fbeck, M.~Spanring, D.~Spitzbart, W.~Waltenberger, J.~Wittmann, C.-E.~Wulz\cmsAuthorMark{1}, M.~Zarucki
% ...
\vskip\cmsinstskip
% ...
1:~~Also at Vienna University of Technology, Vienna, Austria\\
% ...

\end{sloppypar}
\end{document}